\documentclass[%
 reprint,
 amsmath,amssymb,
 aps,
 pra,
]{revtex4-2}

\usepackage{graphicx}
\usepackage{dcolumn}
\usepackage{bm}

\usepackage[
text={18.7cm,10in},centering,
]{geometry}
\usepackage{graphicx}       
\usepackage{subcaption}
\usepackage{subfloat}
\usepackage{braket}
\usepackage{bbm}
\usepackage{amsmath}
\usepackage{xcolor}
\usepackage{tabularx}
\usepackage{multirow}
\graphicspath{{media/}}     
\usepackage{comment}
\usepackage{caption}

\begin{document}

\preprint{APS/123-QED}

\title{Reconstructing Quantum States Using Basis-Enhanced Born Machines}

\author{Abigail McClain Gomez}
 \affiliation{Department of Physics, Harvard University \\ Cambridge, Massachusetts 02138, USA}
\author{Kadijeh Najafi}
\affiliation{IBM Quantum, IBM T.J. Watson Research Center \\ Yorktown Heights, NY 10598 USA}%

\author{Susanne F. Yelin}
\affiliation{Department of Physics, Harvard University \\ Cambridge, Massachusetts 02138, USA}%

\date{\today}

\begin{abstract}
Rapid improvement in quantum hardware has opened the door to complex problems, but the precise characterization of quantum systems itself remains a challenge. To address this obstacle, novel tomography schemes have been developed that employ generative machine learning models, enabling quantum state reconstruction from limited classical data. In particular, quantum-inspired Born machines provide a natural way to encode measured data into a model of a quantum state. Born machines have shown great success in learning from classical data; however, the full potential of a Born machine in learning from quantum measurement has thus far been unrealized. To this end, we devise a complex-valued basis-enhanced Born machine and show that it can reconstruct pure quantum states using projective measurements from only two Pauli measurement bases. We implement the basis-enhanced Born machine to learn the ground states across the phase diagram of a 1D chain of Rydberg atoms, reconstructing quantum states deep in ordered phases and even at critical points with quantum fidelities surpassing 99\%. The model accurately predicts quantum correlations and different observables, and system sizes as large as 37 qubits are considered. Quantum states across the phase diagram of a 1D XY spin chain are also successfully reconstructed using this scheme. Our method only requires simple Pauli measurements with a sample complexity that scales quadratically with system size, making it amenable to experimental implementation.

\end{abstract}

\maketitle


\section{\label{sec:intro}Introduction}

Recent progress towards the creation of highly controllable and programmable quantum systems has made it increasingly possible to tackle hard problems in the realm of fundamental science and real world application \cite{IBM_eagle,arute_quantum_2019,IBM2019,Zhong2020,Gong2021,wu2021strong,ebadi2021quantum}. However, characterization of a target quantum state is limited by the information loss inherent to projective measurements and the noise that pervades today's quantum hardware. Quantum state tomography (QST) refers to the task of reconstructing an unknown quantum state $\rho$ from measurement for characterization. Traditional methods of QST suffer from exponential scaling in system size $N$, limiting their applicability to small sizes ($N<10$) which have been outgrown by modern quantum systems \cite{QST1,QST2,QSt3}. Beyond the concern of exponential scaling with $N$, the presence of quantum correlations in a target quantum state can present another serious obstacle to classical methods of QST.

A full state reconstruction is not necessary for many practical applications. Using randomized measurement tools, it is possible to bypass a complete model for $\rho$ and still estimate functions of $\rho$, thereby tremendously reducing the number of experiments and amount of classical post processing required, and potentially side-stepping the issue of exponential scaling \cite{ Randomized_measure1, van_Enk_random, Elben_random, aaronson_shadow_2018, aaronson_gentle_2019}. Classical shadow tomography, which involves constructing a succinct classical representation of the state from measurements, has been successful in the efficient estimation of quantum properties and in classification tasks that involve local observables \cite{aaronson_shadow_2018, aaronson_gentle_2019, huang_predicting_2020, huang_provably_2021, hu_logical_2022}. The scheme has been proven to asymptotically reach the scaling limit imposed by quantum information theory in terms of number of measurements required, but globally entangling gates are necessary to enjoy less-than-exponential scaling for the task of fidelity estimation \cite{huang_predicting_2020}. To tackle tasks like fidelity estimation, alternative recent tomography schemes employ classical generative machine learning (ML) tools like restricted Boltzmann machines (RBMs), recurrent neural networks (RNNs) and attention-based tomography (AQT) \cite{carleo_solving_2017, torlai_integrating_2019, torlai_neural-network_2018, carrasquilla_reconstructing_2019, wang_scalable_2020, carrasquilla_neural_2021, cha_attention-based_2022}. These ML approaches learn directly from raw data but generally require random measurements from an informationally complete (IC) set of positive operator valued measures (POVMs), which unfavorably scales as $4^N$ for an $N$-qubit system and is not required for general pure state reconstruction \cite{finkelstein_pure-state_2004, flammia_minimal_2005, heinosaari_quantum_2013}.

Born machines are quantum-inspired generative models for machine learning that are based on the probabilistic nature of quantum mechanics \cite{han_unsupervised_2018,wang_scalable_2020,Ising_Born,TN_NISQ_2021,MBL_BORN, cheng_information_2018, coyle_born_2020, mcclain_gomez_born_2021}. The goal of a Born machine is to find a set of probability amplitudes such that the associated Born probabilities would represent the total probability distribution of the target data, mirroring the underlying objective of maximum-likelihood approaches to QST. In a simple application of a Born machine, a single probability distribution is learned by providing training data sampled from the target distribution (single-basis learning) \cite{han_unsupervised_2018, stoudenmire_supervised_2016, mcclain_gomez_born_2021}. Tensor networks are the primary Born machine ansatz for encoding these probability amplitudes, as their structure can innately capture quantum correlations \cite{han_unsupervised_2018,Born_TTN,TN_NISQ_2021}. 
This is advantageous for quantum applications as it is generally known that quantum correlations can make the probability distributions obtained from a quantum state more difficult to learn classically, yet by the same token are one of the main prospects for quantum advantage \cite{arute_quantum_2019,anders_computational_2009, buhrman_nonlocality_2010, liu_entanglement-based_2021}. In fact, it is possible to harness quantum correlations in generative modeling, thereby increasing the model's expressivity and transporting the algorithm into the realm of quantum machine learning \cite{gao_enhancing_2021,glasser_expressive_2019,Ising_Born}. 

In this work, we extend the capability of the Born machine to accept data from different measurement bases (basis-enhancement) and thereby efficiently achieve pure quantum state reconstruction. Furthermore, in contrast to using a randomized measurement toolbox or an IC set of POVMs, we show that by using a priori knowledge of the operators contained in the Hamiltonian, it is possible to accurately train a tensor network-based Born machine using measurements from only two distinct bases. More specifically, we use Born machines to produce generative models of quantum states across ordered phases and close to critical points of a one-dimensional Rydberg chain. This system is characterized by strong Rydberg-Rydberg interactions which produce a controllable platform for quantum computation \cite{jaksch_fast_2000, bernien_probing_2017, weimer_rydberg_2010}. The system also exhibits unusual non-ergodic dynamics for special states known as quantum scars \cite{turner_weak_2018, heller_bound-state_1984}. Beyond these features, the 1D Rydberg chain manifests a nontrivial ground state phase diagram \cite{rader_floating_2019, keesling_quantum_2019}, with multiple crystalline phases that emerge out of a disordered phase for different regions of system parameters. We further explore the critical points at the boundary of ordered phases to test the Born machine's ability to capture long range ordering. We numerically show that a basis-enhanced Born machine (BEBM) can successfully predict the quantum correlations at the critical point and can learn an unknown quantum state to a quantum fidelity surpassing 99\% across different phases for system sizes up to $N = 37$, outperforming the original Born machine. The 1D anisotropic XY chain is also briefly investigated, and it is found that the basis-enhanced Born machine can capture states from the oscillatory region of the phase diagram where previous applications of the single-basis Born machine were unsuccessful \cite{mcclain_gomez_born_2021}.
\begin{figure}
    \hspace{-3.6cm}
    \begin{minipage}[b]{4.2cm}
    \begin{subfigure}[b]{0.22\textwidth}
        \centering
        \includegraphics[width = 4.2cm]{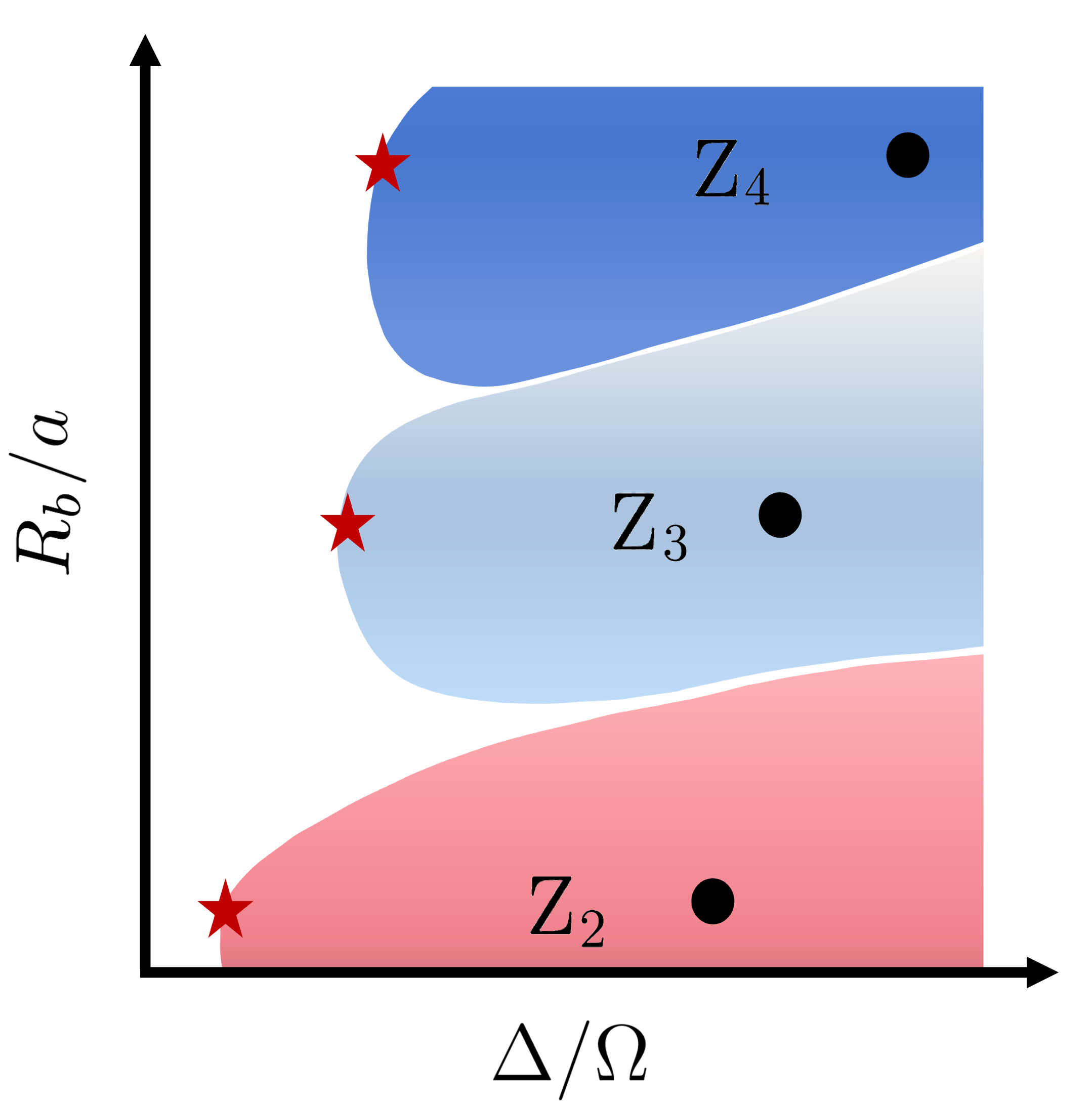}
        \caption{}
        \label{subfig:cartoon-phase-map}
    \end{subfigure}  
    \end{minipage}
    \begin{minipage}[b]{4.2cm}
    \begin{subfigure}[b]{0.23\textwidth}
        \centering
        \includegraphics[width = 4.3cm]{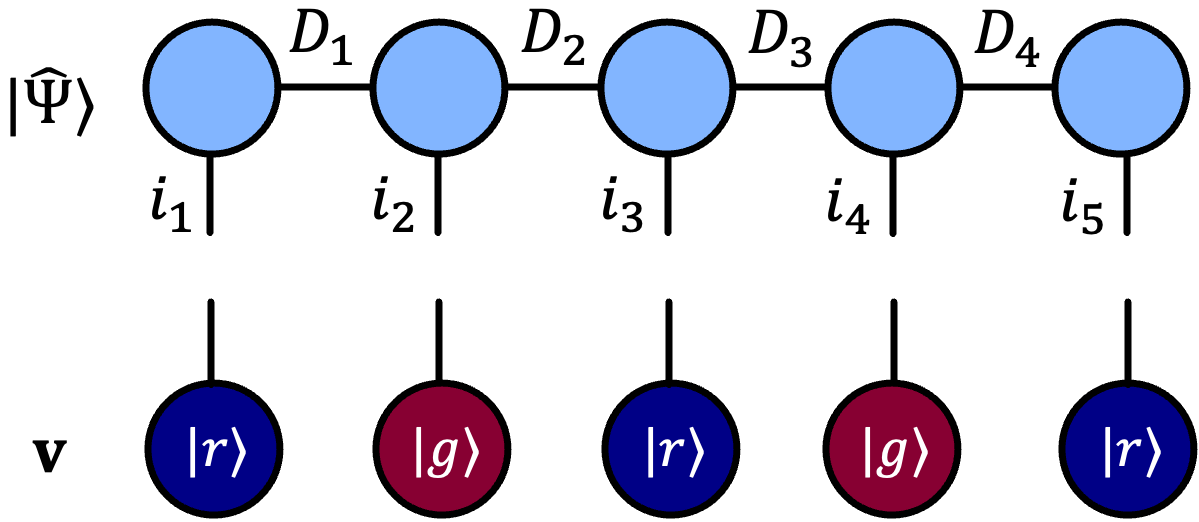}
        \caption{}
        \label{subfig:MPS}
    \end{subfigure}
    
    \begin{subfigure}[b]{0.23\textwidth}
        \centering
        \includegraphics[width = 4.3cm]{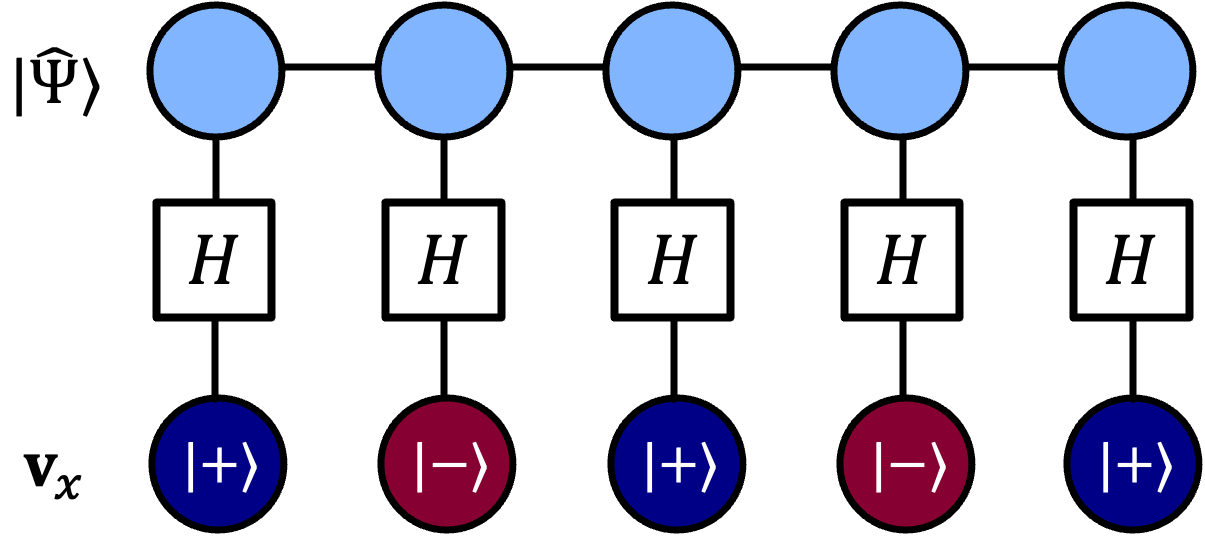}
        \caption{}
        \label{subfig:xMPS}
    \end{subfigure}
    \end{minipage}
    \caption{\ref{subfig:cartoon-phase-map} Schematic picture of the ground state phase diagram for a 1D Rydberg system, indicating three emergent ordered phases as function of detuning and Rydberg radius. The critical points (red stars) and points of interest in ordered phases (black circles) are highlighted. \ref{subfig:MPS} Tensor diagram illustrating how data native to the model can be accepted for learning. \ref{subfig:xMPS} Tensor diagram illustrating how the model can be rotated into another basis to accept data in a non-native basis.}
    \label{fig:phasemap-and-tensors}
\end{figure}

\begin{figure*}[ht]
    \centering
    \includegraphics[width=17.8cm]{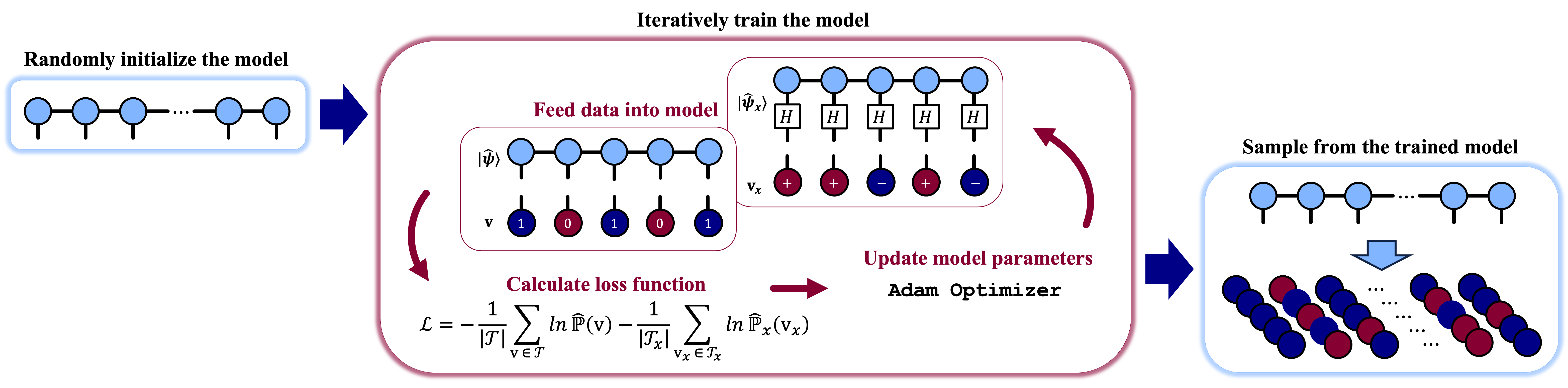}
    
    \caption{Schematic diagram of the training procedure for a basis-enhanced Born machine with data from the $z$- and $x$-bases. First, a chain of tensors forming the MPS ansatz are randomly initialized. The model is then iteratively trained by feeding a batch of measurement data into the model, calculating the loss function, and updating the model parameters through stochastic gradient descent optimizer such as Adam. After training, the model is sampled to produce new configurations of the system.}
    \label{fig:overview}
\end{figure*}

\section{The 1D Rydberg Model}
\label{sec:rydchain}
The focus of this work is a 1D chain of Rydberg atoms. A single Rydberg atom can be treated as a two level system with ground state $\ket{g_i}$ and Rydberg excited state $\ket{r_i}$. Two excited Rydberg atoms in close proximity to each other will interact via the Van der Waals interaction, which must be included in the Hamiltonian of a chain of Rydberg atoms:
\begin{equation}
    \frac{\mathcal{H}}{\hbar} = \sum_i \frac{\Omega_{i}}{2} \sigma_{x}^{i} - \sum_{i} \Delta_{i} n_{i} + \sum_{i<j} V_{ij} n_{i} n_{j}.
    \label{eq:Ham}
\end{equation}
Here, $\sigma_{x}^{i} = \ket{g_i} \bra{r_i} + \ket{r_i} \bra{g_i}$, $n_i = \ket{r_i} \bra{r_i}$, $V_{ij} = C_{6}/r^{6}_{ij}$ represents the Van der Waals interaction, and $\Omega_i$, $\Delta_i$ represent the Rabi frequency and detuning at the $i$\textsuperscript{th} site. Note that in terms of Pauli operators, $n_i =\frac{1}{2}((\sigma_{z}^{i})^2 - \sigma_{z}^{i})$. For simplicity, throughout this work we consider a homogeneous Rabi frequency and detuning ($\Omega_i = \Omega$, $\Delta_i = \Delta$), as well as an equidistant chain geometry where all atoms are separated from their neighbors by a distance $a$.

Due to the Van der Waals interaction, an ensemble of Rydberg atoms resonantly coupled to a single Rabi drive can only sustain one Rydberg excitation within the Rydberg radius ($R_b$), which is defined by equating the relevant energy scales of the system ($\Omega = V_{ij}(r = R_b)$). This blockade mechanism leads to ordered phases in the ground state phase diagram of a 1D Rydberg chain (Fig. \ref{subfig:cartoon-phase-map}) \cite{bernien_probing_2017}. The ordered phases arise out of disorder as the detuning is increased, and higher order phases can be reached by decreasing the inter-atomic distance.

Throughout this work, a Rabi frequency $\Omega$ of 2 MHz and a Van der Waals coefficient $C_6$ of 5.4e-24 m\textsuperscript{6}/s were used for all computations, thereby fixing the Rydberg radius; these numerical values are taken from \cite{bernien_probing_2017}. A more precise phase map is available in the SI Appendix (Fig. S1), produced numerically using ITensor's density-matrix renormalization group (DMRG) algorithm \cite{schollwock_density-matrix_2011,fishman_itensor_2020, white_density_1992, stoudenmire_studying_2012}. The long-range interaction in the Hamiltonian was truncated in calculations to include at most the fifth nearest neighbor. In order to converge on a solution in the ordered side of the phase diagram, calculations were carried out from left to right along a particular $R_b/a$ line. Each new solution was used as the initial guess for a slightly larger $\Delta / \Omega$, a procedure that closely resembles an experimental adiabatic approach. We further treat the results of the DMRG algorithm as ground truth for the purposes of quantifying how well a model can learn a particular quantum state.

Points of interest across the phase diagram were chosen along lines of fixed $R_b/a$ in each ordered phase. Specifically, the critical point at the boundary of the ordered phase and one point deep within the ordered phase were selected for further study; these points are highlighted in Fig. \ref{subfig:cartoon-phase-map}. The atomic separation associated with each ordered phase was taken from \cite{bernien_probing_2017}, with the exception of $Z_4$. A slightly smaller atomic separation is used in this work for the $Z_4$ phase to avoid crossing the $Z_3$-$Z_4$ phase boundary at large detuning $\Delta / \Omega$. For the $Z_2$, $Z_3$, and $Z_4$ ordered phases, the critical point along a fixed $R_b/a$ line was determined by examining three relevant parameters: the bipartite entanglement entropy, the magnetization, and the first excited energy gap \cite{samajdar_complex_2020}. More details can be found in the SI Appendix.

\section{The Generative Born Machine}
Generative models are a central paradigm in unsupervised learning, aiming to capture the total joint probability distribution of the data. This is in contrast to discriminative models, which try to estimate the conditional probabilities often for classification tasks \cite{Generative_review}. Once training is performed, a generative model can be used to produce new data according to the modeled probability distribution -- hence its name. Inspired by the probabilistic nature of quantum mechanics, Born Machines model the quantum wave function $\Psi(\mathbf{v})$ (or more generally, the \textit{probability amplitude} distribution) rather than directly modeling the target distribution $\mathbbm{P}(\mathbf{v})$. The two quantities are related by Born's rule: 
%
\begin{equation}
    \hat{\mathbbm{P}}(\mathbf{v}) = \frac{\lvert \hat{\Psi}(\mathbf{v})\rvert^2}{\sum_{\mathbf{v}\in \mathcal{V}}{\lvert \hat{\Psi}(\mathbf{v})\rvert^2}}.
    \label{eq:Borns Rule}
\end{equation}
To differentiate the model from the target quantum state $\Psi(\mathbf{v})$, we will denote our trained model with a hat: $\hat{\Psi}(\mathbf{v})$. 


While representing a generic probability distribution requires a number of parameters exponential in the system size, it turns out one can harness the underlying structure in natural data to approximate $\mathbbm{P}(\mathbf{v})$ with only a polynomial number of parameters. 
This has been particularly successful in probabilistic graphical models (PGM) \cite{cheng_information_2018,Lin2017-yi}. By a similar analogy, a tensor network can provide an efficient representation of quantum many body states by applying low rank decomposition \cite{TN_efficient,Poulin2011}. A particular factorization that has shown great success in representing a wide category of quantum states is the matrix product state (MPS), which is a one-dimensional tensor network (TN). An MPS is an efficient way of classically storing information from a high-dimensional (Hilbert) space with a lower-dimensional representation, when the data satisfies the area law for entanglement \cite{hastings_area_2007, eisert_colloquium_2010, lu_tensor_2021}. For a one-dimensional chain of atoms such as our system of interest, the mapping is well-established, and the number of parameters scales polynomially with system size \cite{baumgratz_scalable_2013, lanyon_efficient_2017, perez-garcia_matrix_2007, huggins_towards_2019}.
Taking advantage of these facts, we use an MPS as the model ansatz with one tensor for each atomic site: 
\begin{equation}
 \label{eq:MPS}
{\hat{\Psi}} = \sum\limits_{D_1 = 1}^{D}\cdots \sum\limits_{D_{N-1} = 1}^{D}{ M_{1,i_1}^{D_1}\, M_{2,i_2}^{D_1,D_2}\, \cdots \,M_{N,i_{N}}^{D_{N-1}} }
\end{equation}
where each tensor has one spin index $i$ and is connected to neighboring tensors via bond dimensions $D$. The tensor elements themselves are the model parameters that are optimized during training. For simplicity, we use the same bond dimension $D$ between each tensor and select $D$ prior to learning. 

We train the parameters through the minimization of the negative log likelihood (NLL), defined below:
\begin{equation}
    \mathcal{L} = -\frac{1}{\lvert \mathcal{T} \rvert}\sum_{\mathbf{v} \in \mathcal{T}} \ln \hat{\mathbbm{P}}(\mathbf{v}).
    \label{eq:OG Loss}
\end{equation}
In the above expression, $\hat{\mathbbm{P}}(\mathbf{v})$ is the probability distribution of our model, defined in Eq. \ref{eq:Borns Rule}, and $\mathcal{T}$ represents the set of training data. Minimization of the NLL will maximize the model's probability of elements $\mathbf{v}$ that appear in the training data. If the probability distribution $\hat{\mathbbm{P}}(\mathbf{v})$ perfectly captures the probability distribution of the training data, then the NLL will approach the Shannon entropy of the training data $S(\mathcal{T}) = -\sum_{\mathbf{v}} \mathcal{T}(\mathbf{v}) \ln(\mathcal{T}(\mathbf{v}))$, where $\mathcal{T}(\mathbf{v})$ represents the probability of configuration $\mathbf{v}$ within the data $\mathcal{T}$. To accurately capturing the quantum state of interest, it is essential to determine the required size of $\mathcal{T}$. We used a Monte Carlo based-approach focusing on both local and non-local observables: magnetization and Renyi entropy. The details of this calculation can be found in the SI Appendix. 

\subsection{Basis Enhancement}
In a traditional application of a Born machine, the model remains in a fixed basis during learning, and all training data is provided in the same fixed basis \cite{han_unsupervised_2018, mcclain_gomez_born_2021}. We refer to this method of training as learning from a single basis (SB). Note that feeding single-basis data into the model is as simple as projecting a particular configuration $\mathbf{v}$ onto the spin indices of $\hat{\Psi}$. In the language of tensor diagrams, if blue circles represent 1 in the computational basis (or equivalently represent a Rydberg excitation) and red circles represent 0, then feeding a snapshot in the $Z_2$ phase into $\hat{\Psi}$ could be represented as Fig. \ref{subfig:MPS}. Although having complex model parameters has been proven to increase the model expressivity, data from a single basis cannot constrain the relative phases of the probability amplitudes, so purely real parameters are sufficient for SB learning \cite{glasser_expressive_2019}.

\begin{figure*}[ht]
    \centering
    \begin{subfigure}[b]{0.32\textwidth}
        \centering
        \includegraphics[width = 5.7cm]{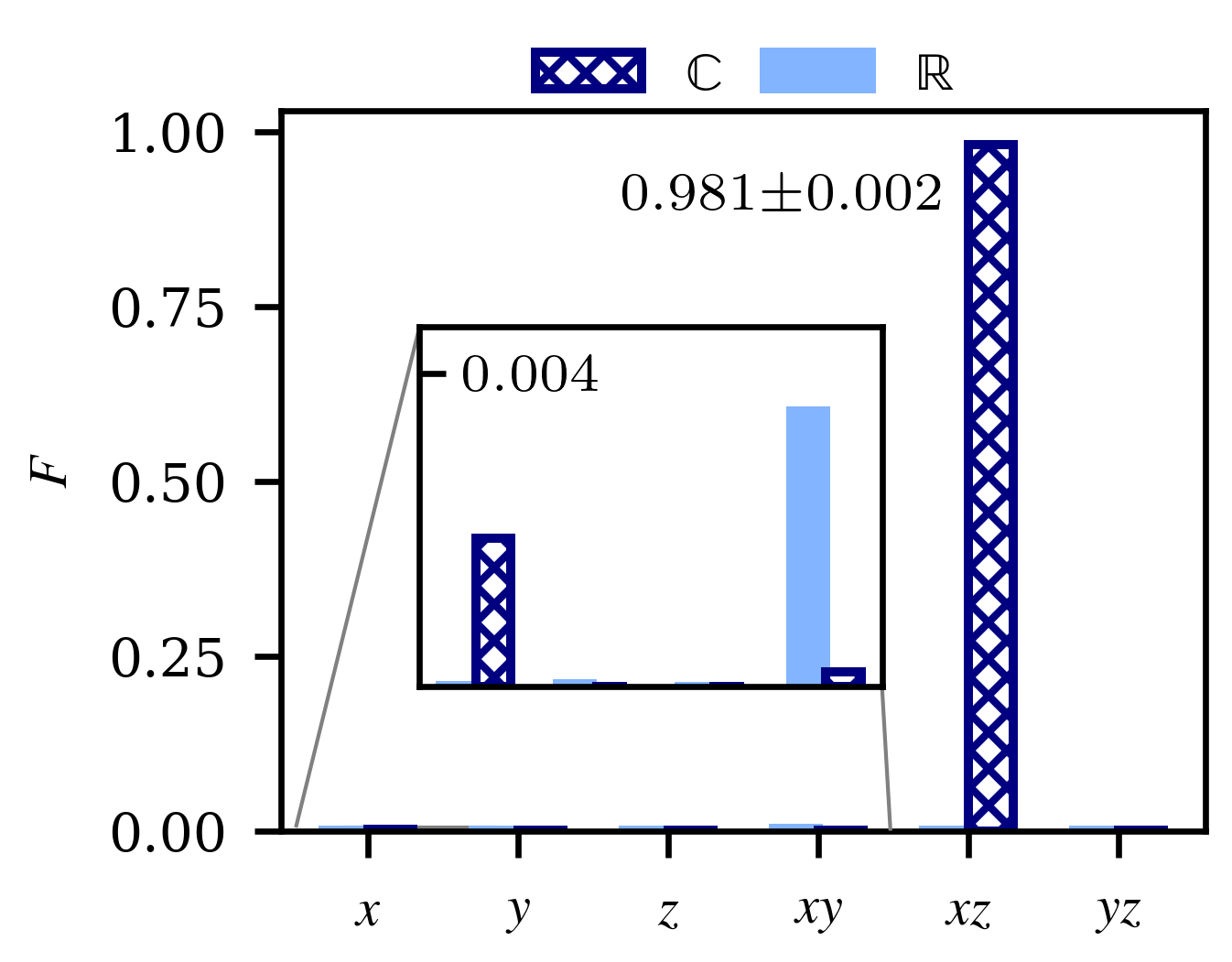}
        \caption{}
        \label{subfig:fids-Z2CP}
    \end{subfigure}
        \hfill
    \begin{subfigure}[b]{0.32\textwidth}
        \centering
        \includegraphics[width = 5.7cm]{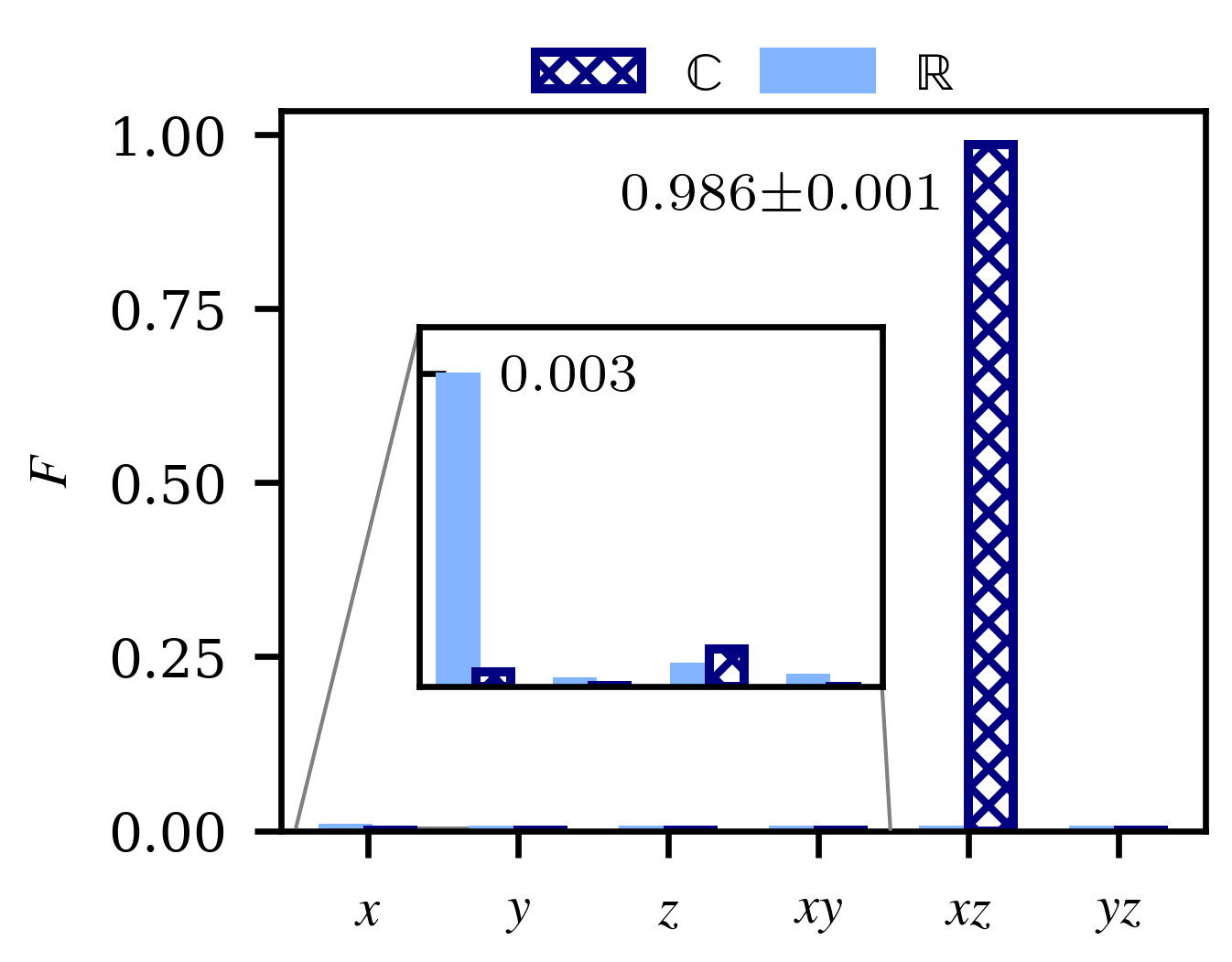}
        \caption{}
        \label{subfig:fids-Z3CP}
    \end{subfigure}
        \hfill
    \begin{subfigure}[b]{0.32\textwidth}
        \centering
        \includegraphics[width = 5.7cm]{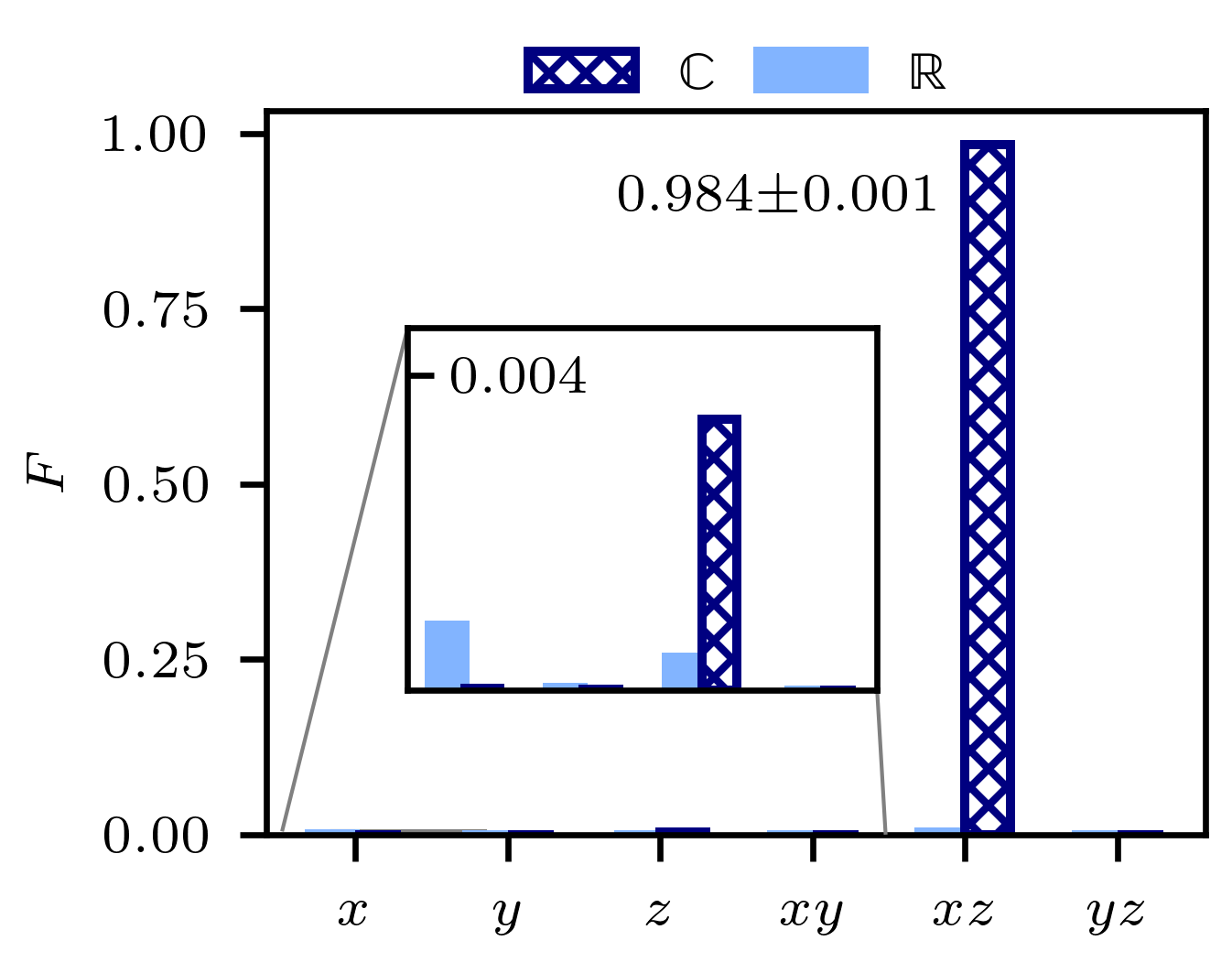}
        \caption{}
        \label{subfig:fids-Z4CP}
    \end{subfigure}
    
    \begin{subfigure}[b]{0.32\textwidth}
        \centering
        \includegraphics[width = 5.7cm]{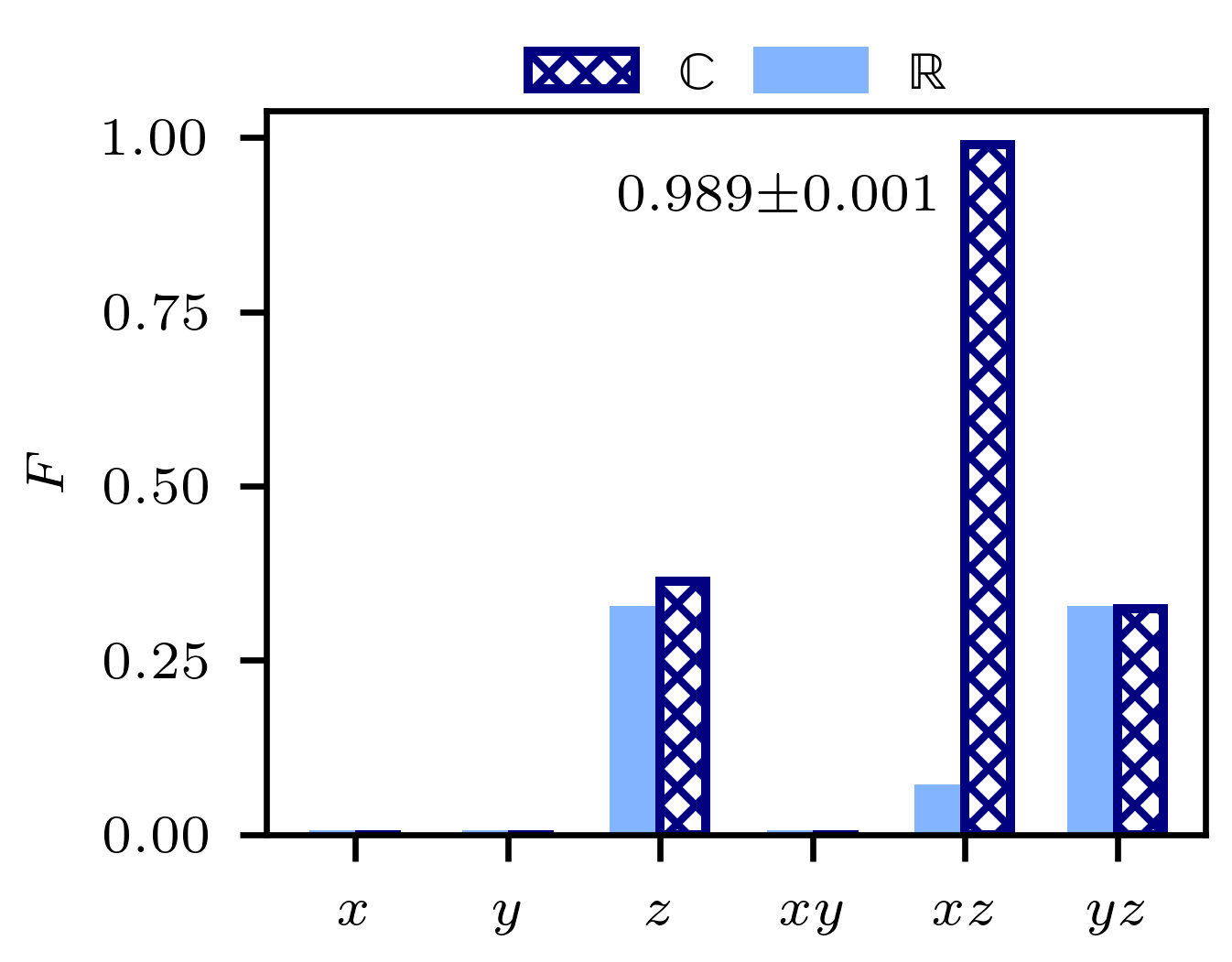}
        \caption{}
        \label{subfig:fids-Z2PH}
    \end{subfigure}
        \hfill
    \begin{subfigure}[b]{0.32\textwidth}
        \centering
        \includegraphics[width = 5.7cm]{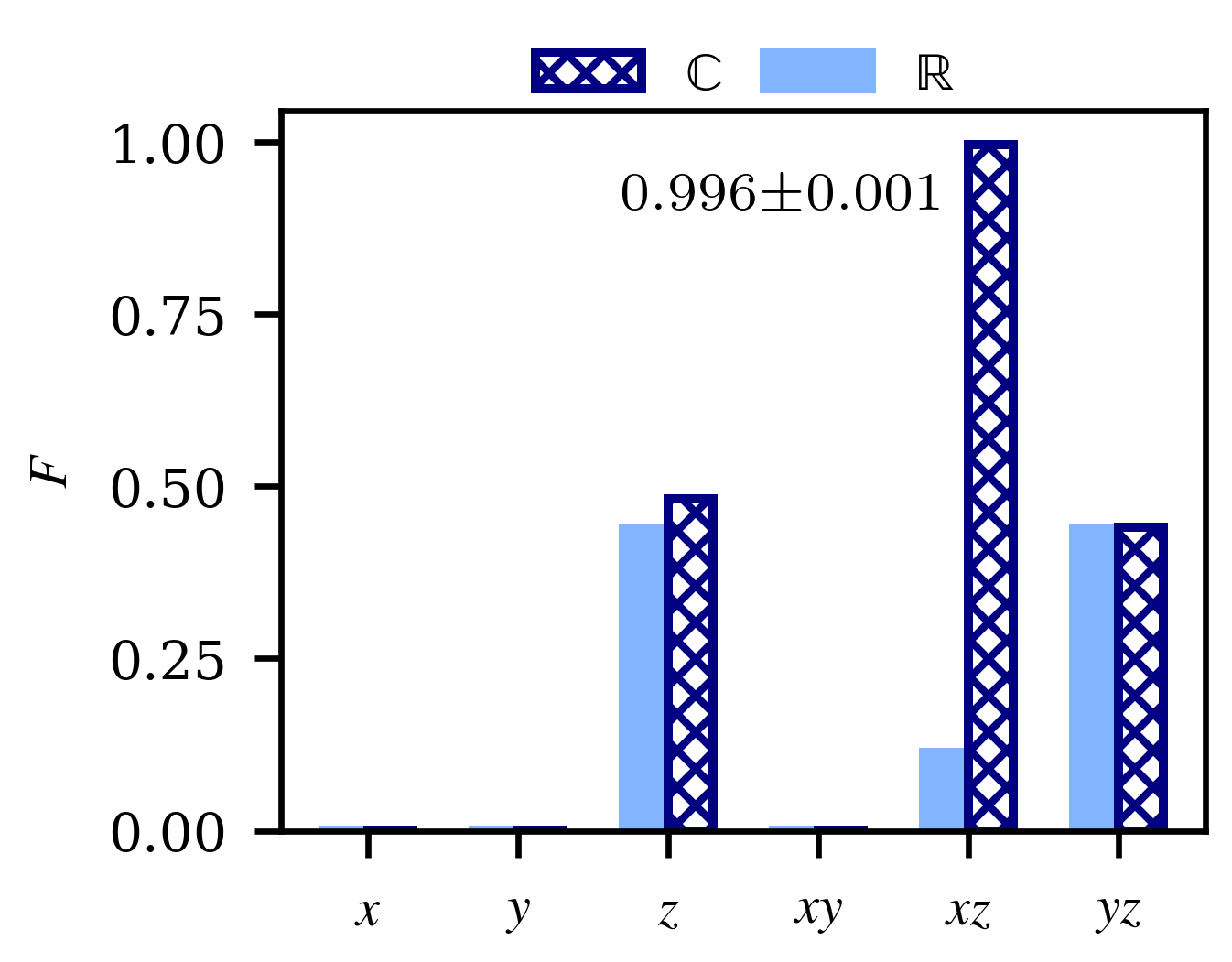}
        \caption{}
        \label{subfig:fids-Z3PH}
    \end{subfigure}
        \hfill
    \begin{subfigure}[b]{0.32\textwidth}
        \centering
        \includegraphics[width = 5.7cm]{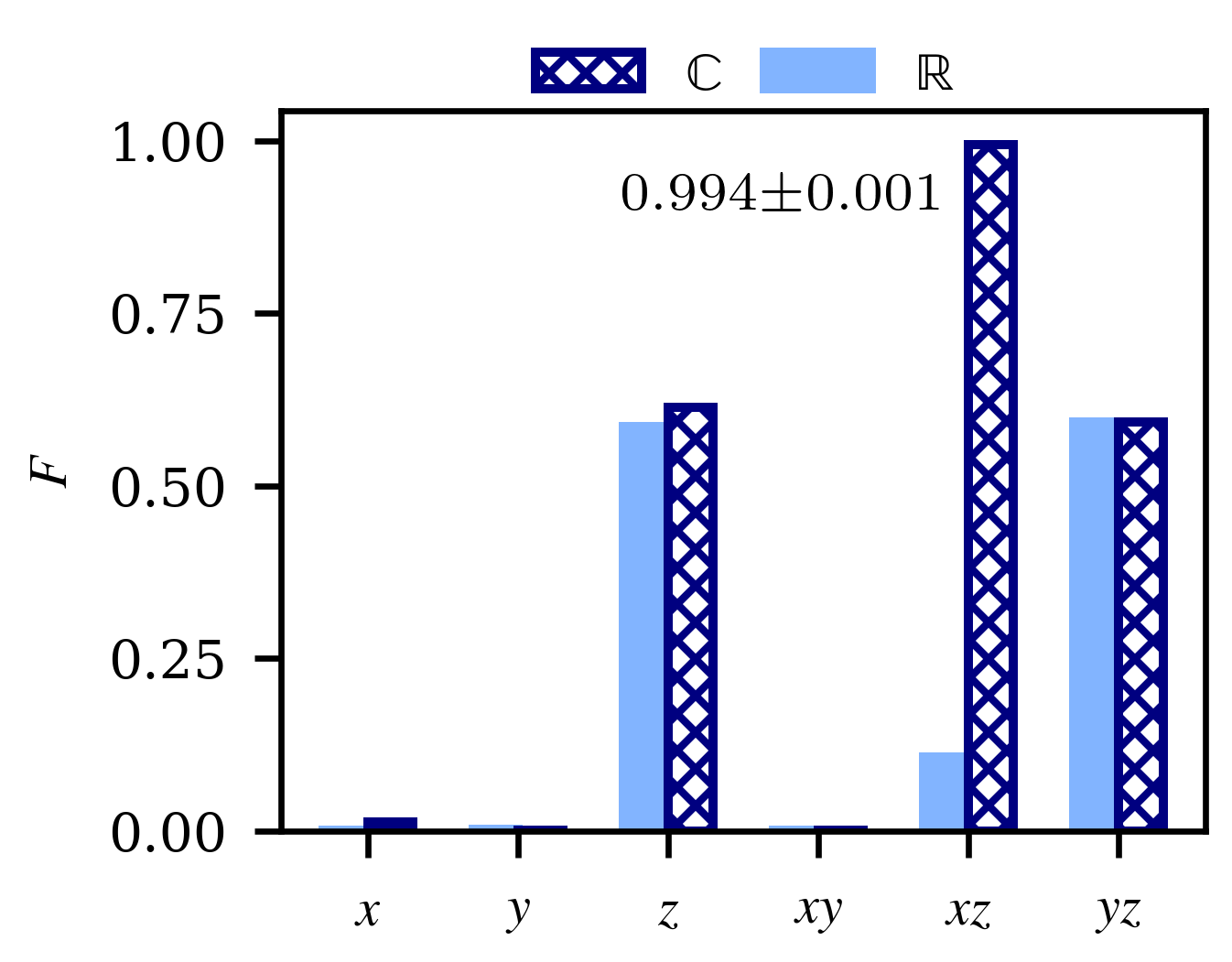}
        \caption{}
        \label{subfig:fids-Z4PH}
    \end{subfigure}
\caption{Fidelity comparison for all combinations of learning basis ($x$-axis) and model parameters (real or complex) for: \ref{subfig:fids-Z2CP} the $Z_2$ critical point, \ref{subfig:fids-Z3CP} the $Z_3$ critical point,  \ref{subfig:fids-Z4CP} the $Z_4$ critical point, \ref{subfig:fids-Z2PH} within the $Z_2$ phase, \ref{subfig:fids-Z3PH} within the $Z_3$ phase, and \ref{subfig:fids-Z4PH} within the $Z_4$ phase. The fidelity of successful choice ($xz$ learning with complex model parameters) is provided, with error given by the standard deviation of the $F$ over the final epoch of training. }\label{fig:fidelities}
\end{figure*}

SB learning can be successful in capturing classical data sets, but it is found wanting for the task of faithful quantum state reconstruction. A Born machine can be modified to allow multi-basis learning that captures more complex data sets; this was alluded to in \cite{wang_scalable_2020, mcclain_gomez_born_2021}, and a version named the Quantum Circuit Born Machine (QCBM) was recently implemented to learn the classical MNIST data set in \cite{rudolph_generation_2021}. Multi-basis quantum machine learning algorithms usually involve random measurements from an IC set of POVMs, but this is not required for the reconstruction of pure states, which are uniquely defined with fewer parameters. Quantum systems are subject to the curse of dimensionality as $2^N$ probability amplitudes are required to define a pure quantum state of $N$ qubits. These amplitudes are generally complex, so $2\times 2^N$ parameters are required (neglecting the normalization constraint and the redundance of global phase) -- see \cite{finkelstein_pure-state_2004, flammia_minimal_2005, heinosaari_quantum_2013} for more rigorous discussions. Global measurements of the system in a single basis will only provide $2^N$ potential outcomes; however, the $2\times 2^N$ constraints necessary to fully define a pure quantum state can be supplied by providing training data from two distinct bases. We chose to maintain the $z$-basis as the initial reference basis for $\hat{\Psi}$. To feed training data from a different basis into the model, $\hat{\Psi}$ must be rotated into the appropriate basis. The MPS architecture can easily be rotated to a non-native basis to accept or generate data in a different basis; this is demonstrated for the case of feeding $x$-basis training data by the application of Hadamard gates $H$ in Fig. \ref{subfig:xMPS}. The loss function must be modified to reflect the basis-enhanced training data. For the case of $z$- and $x$-basis data, $\mathcal{L}$ would take the form: 
\begin{equation}
    \mathcal{L} = -\frac{1}{\lvert \mathcal{T}_{z} \rvert}\sum_{\mathbf{v}_{z} \in \mathcal{T}_{z}} \ln \hat{\mathbbm{P}}_{z}(\mathbf{v}_{z}) - \frac{1}{\lvert \mathcal{T}_{x} \rvert}\sum_{\mathbf{v}_{x} \in \mathcal{T}_{x}} \ln \hat{\mathbbm{P}}_{x}(\mathbf{v}_{x})
    \label{eq:Loss}.
\end{equation} 
In this expression, $\hat{\mathbbm{P}}_{z}$ is the familiar probability distribution defined by the model $\ket{\hat{\Psi}}$, and $\hat{\mathbbm{P}}_{x}$ is the probability distribution defined by the model in the $x$-basis which can be accessed through the application of Hadamard gates ($H^{\otimes n}\ket{\hat{\Psi}}$). In this work, the size of the $x$-basis and $z$-basis data sets are chosen to be the same ($\lvert \mathcal{T}_{z} \rvert = \lvert \mathcal{T}_{x} \rvert$) in order to avoid bias toward one particular basis. Note that the theoretical lower bound of the updated loss is now the sum of the Shannon entropies of the $z$- and $x$-basis training data. We refer to a Born machine that is trained from two measurement bases as a basis-enhanced Born machine (BEBM). A schematic overview of the training process for a BEBM can be found in Fig. \ref{fig:overview}.

Motivated by the fact that complex numbers are required to properly describe quantum systems, we also extend our BEBM to employ complex valued tensors \cite{renou_quantum_2021, ming-cheng_ruling_2022}. Not only do complex parameters increase the expressibility of the model \cite{glasser_expressive_2019}, they also greatly increase the learnability by expanding the number of possible solutions that differ only by global phase. Having a multitude of solutions is particularly advantageous for basis-enhanced learning, as this involves optimizing multiple potentially competing probability distributions with a single model. In our computations, we found that when parameters are restricted to be real, the success of the BEBM is heavily initialization-dependent; however, permitting the parameters to be complex -- providing a multitude of solutions that are identical in practice -- allows the model optimization to evade initialization-dependent behavior. 


\section{Learning across quantum ordered phases via BEBM}
\subsection{1D Rydberg Chain}
\begin{figure*}[ht]

    \centering
    \begin{subfigure}[b]{1.0\textwidth}
    \centering
        \includegraphics[width = 17.8cm]{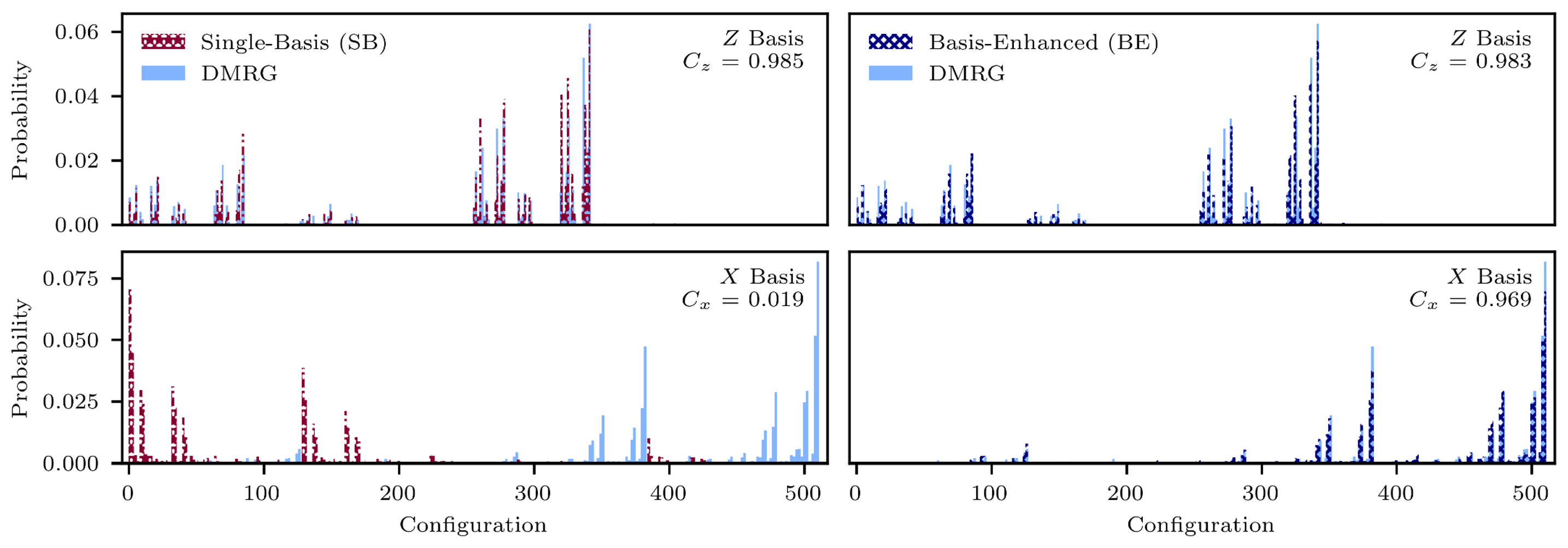}
        \captionsetup{justification=raggedright,singlelinecheck=false}
        \caption{}
    \label{subfig:prob-dist-comp}
    \end{subfigure}
    \begin{subfigure}[b]{1.0\textwidth}
    \centering
        \includegraphics[width = 12.0cm]{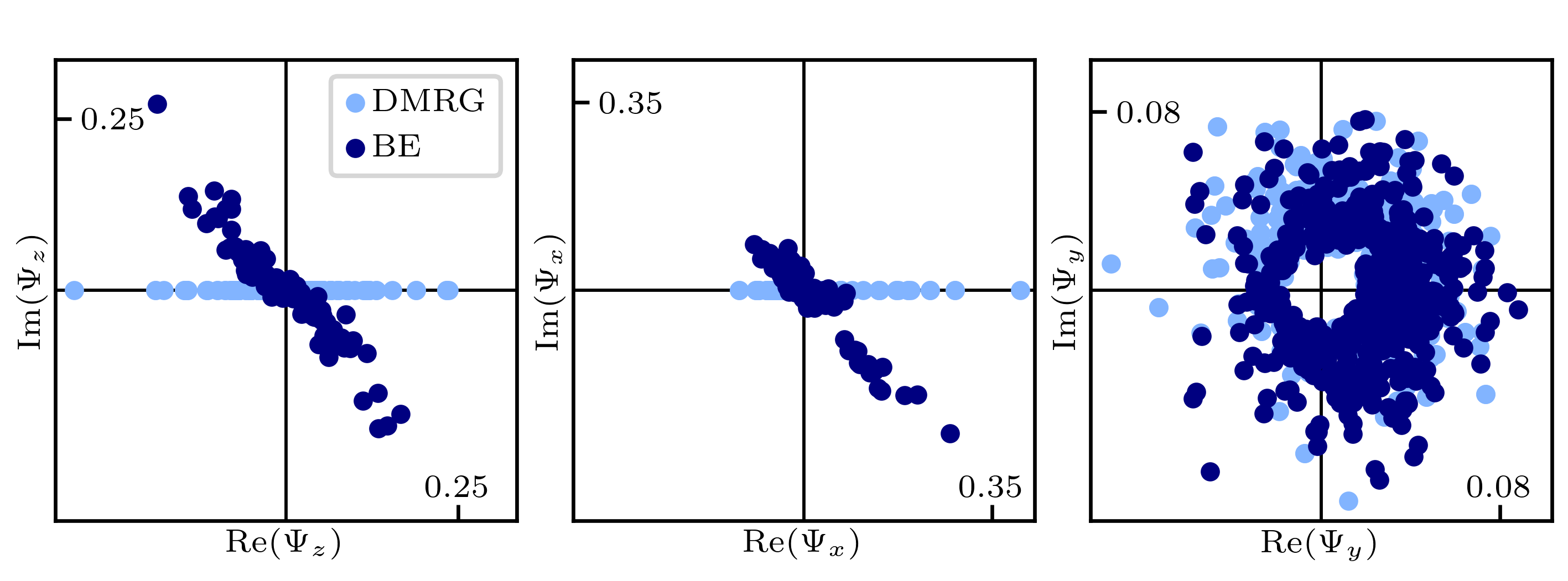}
        \captionsetup{justification=raggedright,singlelinecheck=false}
        \caption{}
        \label{subfig:prob-amps-comp}
    \end{subfigure} 
\caption{\ref{subfig:prob-dist-comp} Overlaid probability distributions in the $z$- and $x$-bases of the DMRG results and: the real-valued, single-basis ($z$) Born machine (left); the complex-valued, basis-enhanced ($xz$) Born machine (right). A smaller chain ($N = 9$) was used for figure clarity, and each histogram bin contains two configurations. The classical fidelities $C_i$ are provided for each basis $i$. \ref{subfig:prob-amps-comp} Overlaid probability amplitudes of the DMRG and complex-valued basis-enhanced Born machine for $xz$ learning of the $Z_2$ critical point. }\label{fig:prob-dist-amp}
\end{figure*}
As a first example, we examine the power of the BEBM in reconstructing the ground states across different ordered phases and at critical points of a one dimensional Rydberg system. By varying the values of the detuning parameter $\Delta$ and the inter-atomic separation $a$, one can cross multiple quantum phase transitions into ordered phases with underlying emergent $Z_2$, $Z_3$ and $Z_4$ symmetries \cite{bernien_probing_2017}. We tasked a variety of basis-enhanced and single-basis Born machines with faithfully reconstructing the quantum state at each point of interest indicated by the dark circles and red stars in the ordered phases and at critical points as shown in Fig. \ref{subfig:cartoon-phase-map}, for Rydberg chain of lengths ranging from $N = 5$ to $N = 37$. The procedure begins by sampling the MPS output of the DMRG algorithm in one or two bases at the parameters of a point of interest, simulating measurement of the physical quantum system and forming the training data set, $\mathcal{T}$. A Born machine is initialized with $N$ tensors (or $2N$ tensors for complex models) and some fixed bond dimension $D$; bond dimensions ranging from $D = 2$ to $D = 16$ were used for analysis. During training, the elements of the tensor network model are updated through gradient descent of the NLL (Eq. \ref{eq:OG Loss} or a form of Eq. \ref{eq:Loss}), using PyTorch's Adam optimizer \cite{kingma_adam_2017}. After training, the model can efficiently generate new samples \cite{ferris_perfect_2012, Najafi_GHZ_nonlocal}. The resulting probability distribution can be compared to that of the training data with a classical metric called classical fidelity, $C$: 
\begin{equation}
    C(\mathbbm{P}, \mathbbm{Q}) = \bigg( \sum_{\mathbf{v}} \sqrt{\mathbbm{P}(\mathbf{v}) \mathbbm{Q}(\mathbf{v})} \bigg)^2 ,
\end{equation}
where $\mathbbm{P}(\mathbf{v})$ ($\mathbbm{Q}(\mathbf{v})$) represents the discrete probabilities contained by probability distribution $\mathbbm{P}$ ($\mathbbm{Q}$). The classical fidelity between probability distributions $\mathbbm{P}$ and $\mathbbm{Q}$ is bounded by 0 and 1, and will only equal 1 when $\mathbbm{P}$ and $\mathbbm{Q}$ are identical.
\begin{table}[ht]
    \centering
    \setlength{\tabcolsep}{0.37em}
    \begin{tabular}{ |c|c|c|c|c|c|c|c|}
    
    \hline
    Basis & $\mathbbm{R}$ / $\mathbbm{C}$ & $C_x$ & $C_y$ & $C_z$ &
    $\mathcal{L} - S$ & 
    $F$\\
    \hline
    \hline

    {\multirow{2}{*}{$x$}}
    & $\mathbbm{R}$ & 0.9433 & 0.5588 & 0.1240 & 0.1317 & 0.0000\\
    & $\mathbbm{C}$ & 0.9619 & 0.4263 & 0.0549 & 0.0720 & 0.0022\\
    \hline
    {\multirow{2}{*}{$y$}}
    & $\mathbbm{R}$ & 0.0035 & 0.9068 & 0.4881 & 0.1477 & 0.0000\\
    & $\mathbbm{C}$ & 0.0022 & 0.9063 & 0.2243 & 0.1485 & 0.0000\\
    \hline
    {\multirow{2}{*}{$z$}}
    & $\mathbbm{R}$ & 0.0034 & 0.9049 & 0.9855 & 0.0346 & 0.0000\\
    & $\mathbbm{C}$ & 0.0073 & 0.8268 & 0.9852 & 0.0340 & 0.0000\\
    \hline
    {\multirow{2}{*}{$xy$}}
    & $\mathbbm{R}$ & 0.9249 & 0.8544 & 0.0346 & 0.5459 & 0.0035\\
    & $\mathbbm{C}$ & 0.9567 & 0.9003 & 0.0184 & 0.2598 & 0.0003\\
    \hline
    {\multirow{2}{*}{\colorbox{yellow}{$xz$}}}
    & $\mathbbm{R}$ & 0.6384 & 0.6009 & 0.5499 & 2.6052 & 0.0004\\
    & \colorbox{yellow}{$\mathbbm{C}$} & \colorbox{yellow}{0.9644} & \colorbox{yellow}{0.9027} & \colorbox{yellow}{0.9859} & \colorbox{yellow}{0.0966} & \colorbox{yellow}{0.9831}\\
    \hline
    {\multirow{2}{*}{$yz$}}
    & $\mathbbm{R}$ & 0.0039 & 0.9079 & 0.9852 & 0.1805 & 0.0000\\
    & $\mathbbm{C}$ & 0.0033 & 0.9078 & 0.9856 & 0.1831 & 0.0000\\
    \hline
    \end{tabular}
    \caption{Classical fidelities in the $x$-, $y$-, and $z$-basis between the training data $\mathcal{T}$ and data sampled from the trained model $\hat{\Psi}$ for real and complex model parameters. The ``Basis'' column indicates from which basis(es) the training data originated. The rightmost columns report final loss $\mathcal{L}$ minus the Shannon entropy of $\mathcal{T}$ (summed over appropriate bases, when applicable) and quantum fidelity $F(\Psi,\hat{\Psi})$. These results are for the $Z_2$ critical point, with a fixed bond dimension of $D = 4$.}
    \label{tab:Z2-results}
\end{table}
The efficacy of the trained model itself can be evaluated through the quantum fidelity. Because the ground truth quantum system is simply an MPS calculated using the DMRG algorithm, the quantum fidelity between our model and the state of the quantum system can be calculated exactly: 
\begin{equation}
    F = \lvert \langle \Psi| \hat{\Psi}\rangle \rvert ^2
\end{equation}
Practically, this calculation amounts to contracting the spin indices of the model with those of the DMRG output state and taking the modulus square. If real experimental data was our only resource, a second full state tomography would be required for this calculation, but for now this simple approach enabled by DMRG provides a further test of our model and gives insight into when and why it is possible to learn a quantum state.

Table \ref{tab:Z2-results} provides the classical fidelities between samples taken from the trained model and samples taken from the DMRG algorithm output. All three orthogonal Pauli bases are provided for verification, although only data from the basis(es) indicated in the ``Basis'' column was used for training. Notice that while there is usually high classical fidelity in the basis(es) used for training, suggesting that all Born machines are successful in a classical sense, only in the highlighted training scenario are all fidelities large -- even the quantum fidelity. For a more visual comparison, Fig. \ref{fig:fidelities} provides the quantum fidelity after training each combination of single-basis and basis-enhanced Born machines with real and complex model parameters, for each point of interest across the phase diagram. Notably, one specific choice -- the basis-enhanced Born machine learning from $xz$ data with complex model parameters -- is successful at every point, while all other learning scenarios fail. The success / failure of different training choices is amplified at the critical points, making up the top row of subfigures in Fig. \ref{fig:fidelities}.

Although Fig. \ref{fig:fidelities} provides insight into which training choice can enable quantum state learning, the quantum fidelity metric supplies little information regarding what happens when learning fails. To examine the cases of failure more closely, the left panel of Fig. \ref{subfig:prob-dist-comp} provides the overlaid probability distributions at the $Z_2$ critical point for single-basis learning -- specifically, $\mathcal{T}_z$ was used with real-valued model parameters. Notice that although there is very good overlap between the data sets in the $z$-basis, a large overlap is not guaranteed in the other bases; this is particularly apparent in the $x$-basis. The Born machine is given no information about the correct relative phases of the quantum state, as phase information cannot be extracted from measurement data in a single basis -- such a data set is purely classical, even if the data is acquired from a quantum system. In contrast, the right panel of Fig. \ref{subfig:prob-dist-comp} highlights the success of the complex-valued basis-enhanced Born machine for the judicious basis choice of $xz$. Each sampled probability distribution has significant overlap with the training data, and the quantum fidelity surpasses 98\% with a modest bond dimension of $D = 4$. 

We also examine the actual probability amplitudes predicted by the basis-enhanced model, to satisfy any remaining skeptics of the model's success. In Fig. \ref{subfig:prob-amps-comp}, the probability amplitudes of the DMRG and basis-enhanced results are provided for each basis state comprising the Hilbert space in the $z$-, $x$-, and $y$-bases. Although a point for point comparison is visually unreasonable, it is evident -- especially for the linear $z$- and $x$-bases results -- that the structure of basis-enhanced amplitudes closely matches that of DMRG with a different global phase, which amounts to a rotation of the amplitudes in the complex plane. 

\begin{figure*}[ht]
    \centering

        \includegraphics[width = 17.8cm]{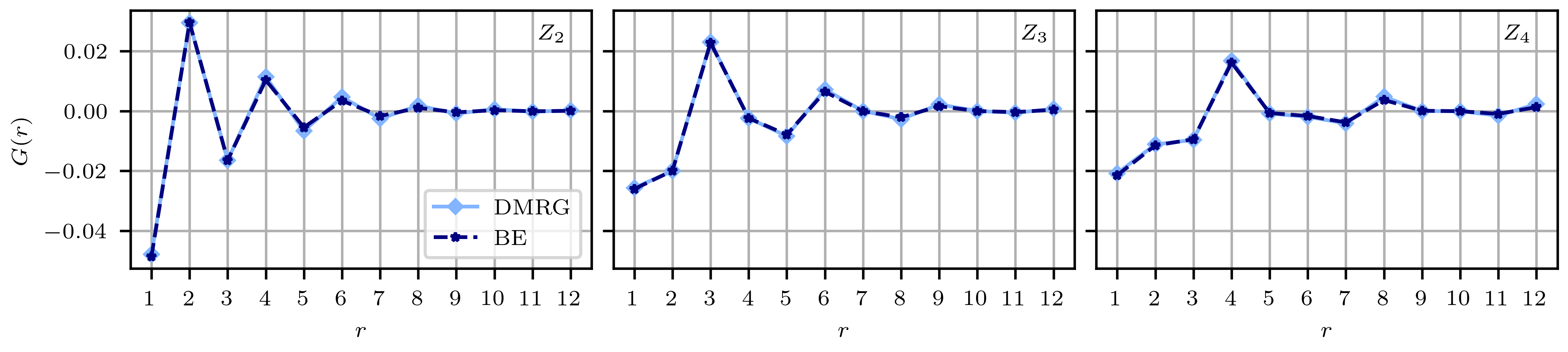}

\caption{Learning correlation function as a function of site distance at critical points across different ordered phases of 1D Rydberg system. For comparison, the results obtained from the BEBM has overlaid with the DMRG results indicating the power of BEBM in capturing quantum correlations at critical point. }
\label{fig:corr-func}
\end{figure*}

It is known that long-range ordering at and near quantum critical points leads to a diverging correlation length. This has been reported to be the primary reason of failure for learning the quantum state close to  a critical point \cite{Torlai_phase_2016, Najafi_Ising_peps}. As a further test of the complex-valued BEBM, we examine the learnabilty of the correlation function close to critical point. The correlation function is defined as: 
\begin{equation}
    G(r) = \sum_{i}\frac{\langle n_{i}n_{i+r} \rangle - \langle n_{i} \rangle \langle n_{i+r} \rangle}{N_{p}},
\end{equation}
in this expression, $n_i$ is the Rydberg state projector at site $i$ as defined in the Hamiltonian, and the normalization $N_{p}$ is the number of pairs of sites included in the sum. $G(r)$ is a measure of Rydberg density correlations separated by a number of sites $r$. There is generally good agreement between $G(r)$ as calculated from the trained model and from the DMRG results (Fig. \ref{fig:corr-func}), indicating the success of the BEBM at capturing quantum correlations at and close to critical point. 

\subsection{1D Anisotropic XY Chain}
To further explore the success and necessity of the basis-enhancement, we applied a Born machine to the anisotropic XY chain with a transverse magnetic field, first attempted in \cite{mcclain_gomez_born_2021}. The system is described by the Hamiltonian: 
\begin{equation}\label{H_XY}\
\mathcal{H}=-J\sum_{i = 1}^{N-1}\Big{(}\frac{1+\gamma}{4}\sigma_i^x\sigma_{i+1}^x+\frac{1-\gamma}{4}\sigma_i^y\sigma_{i+1}^y\Big{)}-\frac{h}{2}\sum_{i = 1}^N\sigma_i^z ,
\end{equation} 
where $J$ is the spin-spin coupling, $\gamma$ captures the relative strength of interaction along $x$ and $y$, and $h$ is the external magnetic field coupled to spins along the $z$-axis; we fixed $J = 1$ as in \cite{mcclain_gomez_born_2021}. Although this system is exactly solvable, it spans multiple interesting phases. In \cite{mcclain_gomez_born_2021}, it was demonstrated that the single-basis Born machine can capture the ground states of the anisotropic XY chain at the Ising critical point ($\gamma = h = 1$), within the ordered phase ($\gamma > 1; h < 1$), and within the disordered phase ($\gamma > 1; h > 1$) with quantum fidelities surpassing 99\% for $z$-basis training data and real model parameters. However, in the oscillatory region ($h^2 + \gamma^2 < 1$) where the system's correlation function exhibits oscillatory behavior, the single-basis Born machine was unsuccessful. 

\begin{figure}[ht]
    \centering
    \includegraphics[width = 8.7cm]{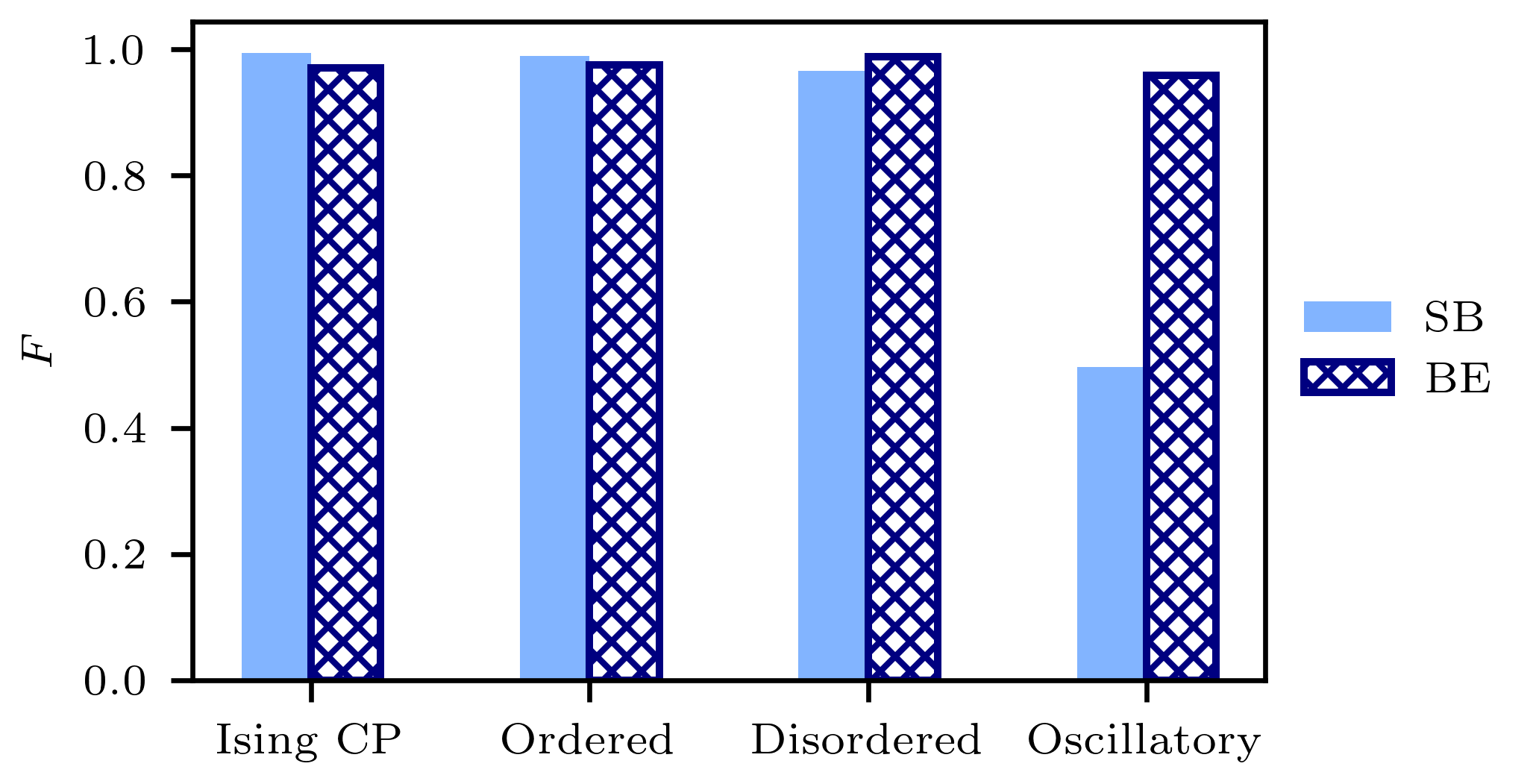}
    \caption{Comparison of quantum fidelity for the real-valued, single-basis ($z$) Born machine and the complex-valued, basis-enhanced ($xz$) Born machine when learning various points from the ground state phase diagram of the 1D anisotropic XY chain. Ising critical point: $h = 1.0, \gamma = 1.0$; ordered: $h = 0.5, \gamma = 1.5$; disordered: $h = 2.0, \gamma = 2.0$; oscillatory: $h = 0.5, \gamma = 0.5$.}
   \label{fig:XY-fids}
\end{figure}

In this work, we applied the single-basis and BEBM to the XY ground states considered by \cite{mcclain_gomez_born_2021}; the results are presented in Fig. \ref{fig:XY-fids}. At fixed bond dimension, the BEBM produces models with quantum fidelities comparable to those of the single-basis Born machine for the states that could be captured via SB learning. Additionally, the BEBM successfully captured the point inside the oscillatory region that could not be learned via SB learning, highlighting the robustness of basis-enhancement.


\subsection{Basis Enhanced Born Machine Under the Hood}
In order to claim knowledge of a full quantum state, it is important to know the relative phases between the probability amplitudes of basis states. Phase information is lost when the probability amplitudes are converted to probabilities, but reconstruction of the lost relative phase information is achievable if one possesses knowledge of the state's probability distribution in multiple bases. This reconstruction is possible because the relative phases in one basis will determine the probability amplitudes of the state -- and hence the probability distribution -- in a different basis. 

A tomographically complete set on the order of $2^{2N}$ measurements is theoretically required to provide enough information to reconstruct an arbitrary quantum state; however, this is an upper bound. Most quantum states of physical interest such as pure states, ground states of local Hamiltonians, and states with particular symmetries can be specified by a small number of parameters and thus for such quantum states, fewer measurements are required (a tomographically \textit{in}complete set) for reconstruction\cite{QST1}. The numerical work presented here illustrates that only measurements in the $z$-basis and $x$-basis are sufficient to reconstruct the ground state of the Rydberg chain. This number of measurements ($2\times 2^N = 16384$ for $N = 13$) is much smaller than what would be considered tomographically complete ($4^N = 67108864$ for $N = 13$). For the XY chain, it was demonstrated that only $z$-basis data is required for state reconstruction in many cases, an even more extreme example of the sufficiency of an incomplete measurement set. 

To illuminate why such limited information is often sufficient for the XY chain, here we investigate the probability amplitudes of the ground states and consider model initialization. 
First, both the Rydberg chain and the XY chain feature real-valued probability amplitudes in the computational basis. This phenomenon can be explained simply by the fact that each system's Hamiltonian is purely real; hence, each Hamiltonian is symmetric rather than being merely Hermitian. Any eigenstate of these Hamiltonians can be written to have purely real probability amplitudes for a particular global phase. In addition to being real, the probability amplitudes of the XY chain ground states are real and non-negative -- see Fig. \ref{fig:XY} in the SI Appendix. This is a feature of some local spin Hamiltonians known as stoquastic Hamiltonians, which are characterized by real and non-positive off-diagonal matrix elements in the standard basis  \cite{bravyi_complexity_2007}. For such Hamiltonians, it is proven that  for any $\beta \geq 0$, the Gibbs density matrix $\rho=e^{-\beta H}/ \,{\rm Tr} \, e^{-\beta H}$ has non-negative matrix elements in the standard basis which simply implies that $|\psi(\textbf{v})\rangle=\sum_{\textbf{v}}\alpha_{\textbf{v}}|\textbf{v}\rangle$ where $\alpha\geq 0$. Thus, the ground state can be fully described by a single probability distribution: $\alpha_{\textbf{v}} = \sqrt{\mathbbm{P}(\textbf{v})}$.  
Moreover, our choice to initialize the real-valued model with random values between zero and one also produces a state with purely positive probability amplitudes. The initial state is therefore close in parameter space to the target ground state of any stoquastic Hamiltonian, and the model arrives at the correct relative phase choice with only $z$-basis training data. Considering the positive nature of the probability amplitudes together with our model initialization explains why single-basis learning can be sufficient for reconstructing the ground states of the XY Hamiltonian in the ordered and disordered phases, as well as at the Ising critical point. We will comment on the failure of single basis learning in the oscillatory phase in the final section. 

Theoretically, quantum state reconstruction of an arbitrary pure quantum state is possible from $2 \times 2^N$ measurements for any basis choice; however, not all probability distributions are created equal when it comes to training a Born machine. It is known that training a Born machine via stochastic gradient descent (SGD) can fail to capture non-local information within a distribution as these methods are based on local updates  \cite{Najafi_GHZ_nonlocal}. We heuristically found that probability distributions which are spread across a large fraction of the basis states are generally more difficult -- or even impossible -- to learn with traditional stochastic gradient descent optimization methods. 
We believe that this is the primary reason learning fails for certain basis choices.  
The degree to which a probability distribution is spread across the $2^N$ configurations can be quantified by its Shannon entropy, which is linked to learning in that at the global minimum of the loss function, $\mathcal{L}$ approaches the Shannon entropy of the training data. A larger Shannon entropy will result in a shallower optimization landscape. 

For each system considered, it was observed that careful selection of measurement bases can improve state reconstruction. In the Rydberg chain results presented in the previous section, the BEBM was successful only when $z$- and $x$-basis training data was employed. When $y$-basis training data was substituted for either, the Born machine was unable to capture the quantum state (see Table \ref{tab:Z2-results} and Fig. \ref{fig:XY-fids}). The importance of basis choice may in part be explained by the operators that appear in the Hamiltonian, as the Rydberg Hamiltonian only involves $\sigma_z$ and $\sigma_x$ operators. This leads to preferential configurations for the ground state in these bases, which in turn produces more structured probability distributions that are easier to learn.
To test this hypothesis, the Rydberg Hamiltonian was modified by replacing the $\sigma_x$ operator with a $\sigma_y$ operator. We found that in this scenario, $z$ and $y$ was the only training data basis combination that resulted in successful learning. Similarly, for the anisotropic XY chain, a second point from the oscillatory region with $\gamma < 0$ was tested. A negative $\gamma$ results the the $\sigma^y$ operators appearing in the Hamiltonian to be weighted more heavily than the $\sigma^x$ operators. We found that for this ground state, $yz$ training data was successful for learning while $xz$ training data failed, in agreement with the hypothesis. See the SI Appendix for more information. These results stress the importance of using a priori knowledge about the Hamiltonian to select the training bases. When this knowledge is not available, an active learning strategy like that employed by \cite{lange_adaptive_2022} could be adapted for Born machines.

\begin{figure}[ht]
        \includegraphics[width=8.7cm]{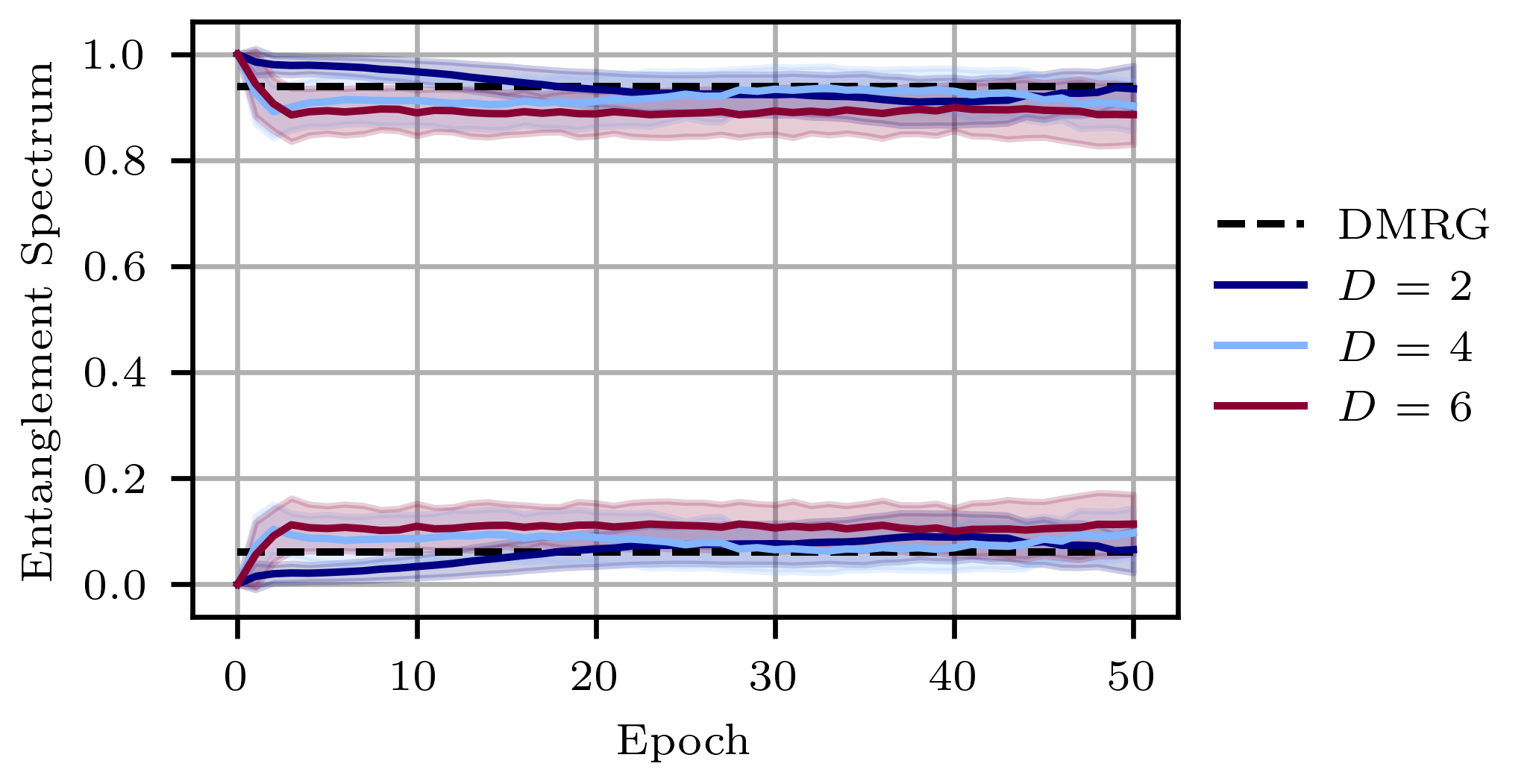}
        \label{fig:ent-spec}
\captionsetup{justification=raggedright,singlelinecheck=false}
\caption{Largest two eigenvalues of the entanglement spectrum for the reduced density matrix of half of the chain, calculated after each epoch for a variety of bond dimension $D$. These results are averaged over 100 trials of learning the $Z_2$ critical point ($N = 13$), for a complex BEBM learning from $x$- and $z$-basis data. The ribbon indicates $\pm$ one standard deviation.}
\label{fig:training}
\end{figure}

Finally, to further investigate the efficiency and robustness of the enhanced-Born Machine, we examined how the entanglement spectrum of the BEBM changes during training at $Z_2$ critical point. Starting from our model MPS, we first obtain the reduced density matrix for half of the quantum spin chain ($N=13$) by tracing out the complementary sites: $\rho_A={\rm Tr}_B (\rho \ln \rho)$, where $\rho=|\hat{\psi}\rangle\langle \hat{\psi}|$ is the density matrix of the full system. Then the entanglement spectrum can be obtained by diagnalizing $\rho_A$. The largest two eigenvalues are plotted in Fig. \ref{fig:training}, as all others were vanishingly small. 
The plot demonstrates that even for modest bond dimensions of $D=2,4,6$, the model can successfully encode the system's entanglement, even at a critical point. In fact, it is even more beneficial to choose a small bond dimension, as over-parameterization can lead to larger oscillations in the spectrum -- see the wider ribbon for the red $D=6$ curve in Fig. \ref{fig:training}.

\subsection{Scaling Trends}

\begin{figure*}[ht]
    \centering
    \begin{minipage}{8.8cm}
    \centering
    \begin{subfigure}[b]{1.0\textwidth}
    \centering
        \includegraphics[width = 8.7cm]{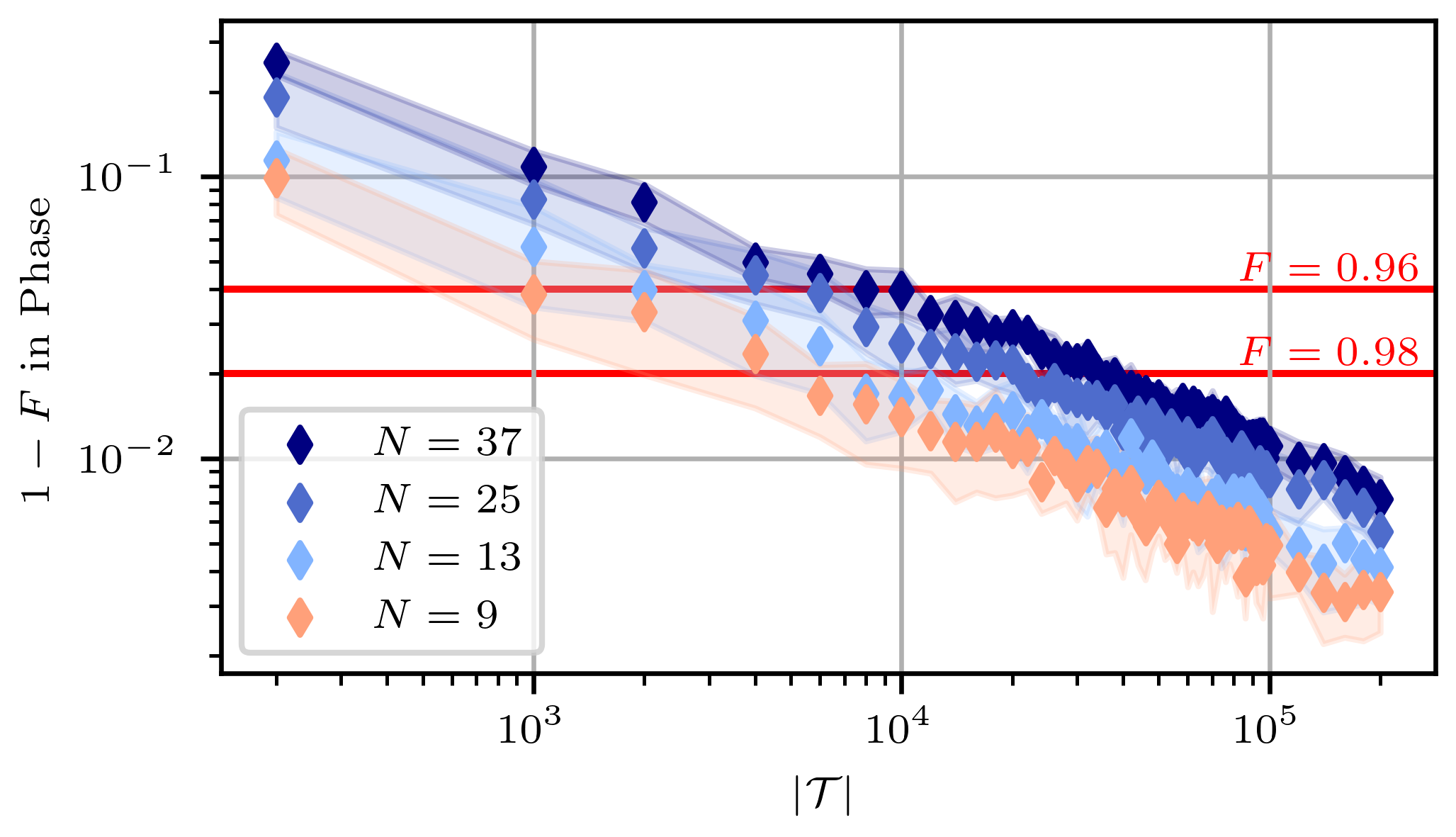}
        \captionsetup{justification=raggedright,singlelinecheck=false}
        \caption{}
        \label{subfig:PH-scaling}
    \end{subfigure} 
    \end{minipage}
    \hfill
    \begin{minipage}{8.8cm}
    \centering
    \begin{subfigure}[b]{1.0\textwidth}
    \centering
        \includegraphics[width=8.7cm]{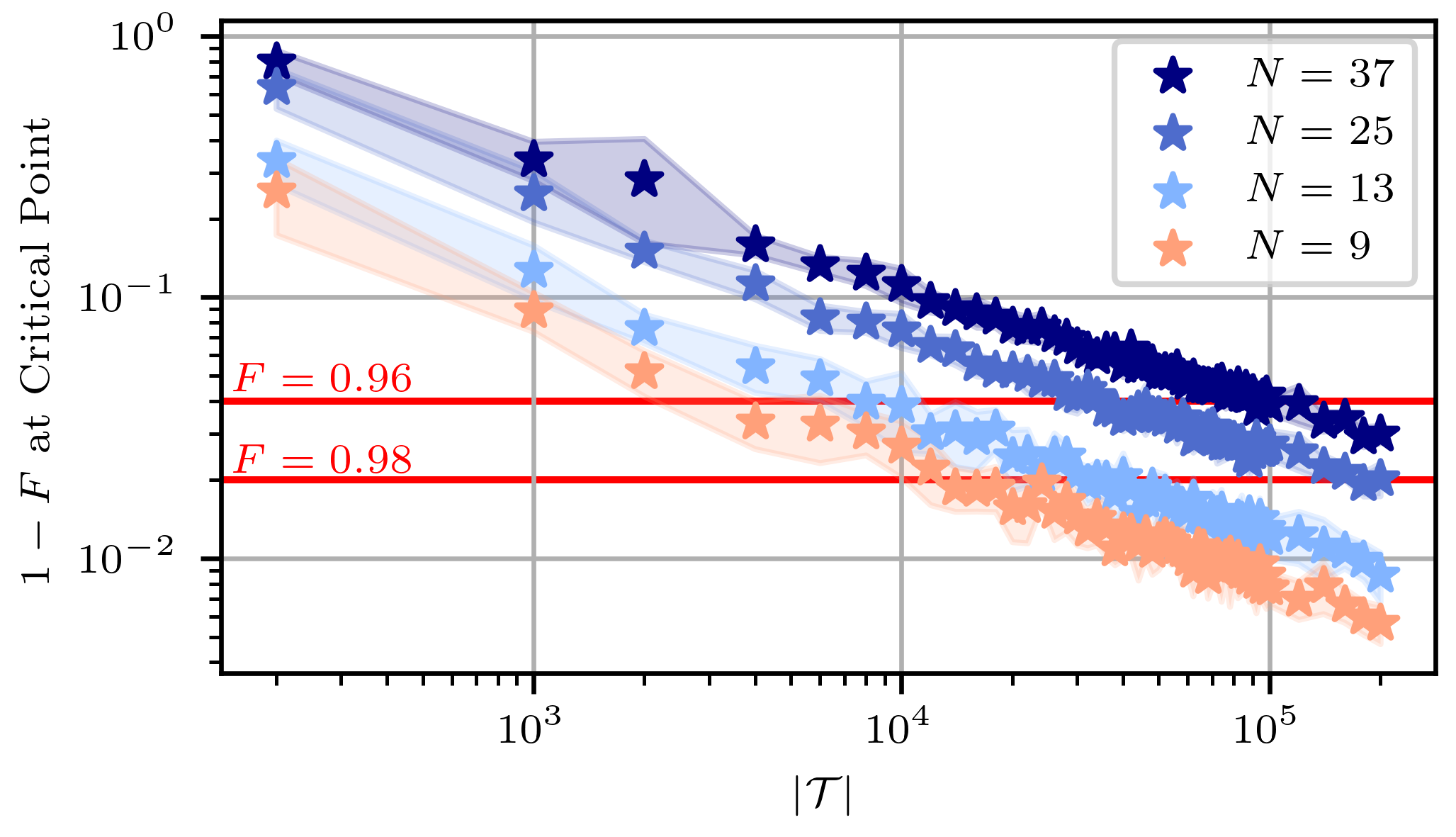}
        \captionsetup{justification=raggedright,singlelinecheck=false}
        \caption{}
        \label{subfig:CP-scaling}
    \end{subfigure} 
    \end{minipage}
    \par\medskip

    \begin{subfigure}[b]{1.0\textwidth}
        \centering
        \includegraphics[width = 17.8cm]{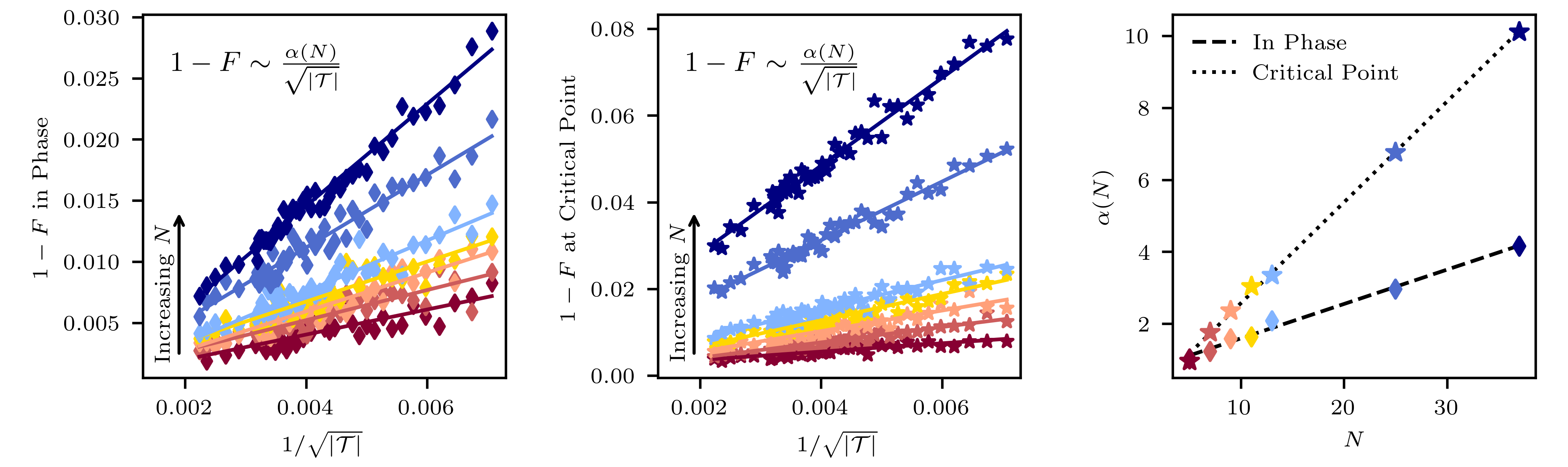}
        \captionsetup{justification=raggedright,singlelinecheck=false}
        \caption{}
        \label{subfig:scaling-trends}
\end{subfigure} 
\caption{\ref{subfig:PH-scaling}: Infidelity after training basis-enhanced born machines within the $Z_2$ phase for a variety of system sizes $N$ as a function of training data size $|\mathcal{T}| = |\mathcal{T}_x| + |\mathcal{T}_z|$. For consistency across different $N$, the value of $\Delta/\Omega$ was selected at fixed $a$ such that the probability of being in the ordered phase is $\sim75\%$. Each point is the average of 10 trials, and the ribbon indicates $\pm 1$ standard deviation of the trials. \ref{subfig:CP-scaling}: Infidelity after training basis-enhanced born machines at the $Z_2$ critical point for a variety of system sizes $N$. \ref{subfig:scaling-trends}: The left panels display the linear relationship between infidelity and $|\mathcal{T}|^{-1/2}$. On the right, the slopes of the linear fits from the left panels are plotted against system size $N$, revealing another linear relationship.}
\label{fig:scaling}
\end{figure*}

To better understand the sample complexity of the BEBM and its scaling with system size $N$, we investigated the accuracy of the BEBM in learning the $Z_2$ critical point as a function of sampling size ($|\mathcal{T}|$) for system sizes ranging from 5 to 37 and fixed bond dimension $D = 4$.
To contrast the complexity of sampling at the critical point, we compare the results with those within the ordered phase.
The results are provided in Fig. \ref{fig:scaling}. Overall, the scaling law $1 - F \sim c \times N \times |\mathcal{T}|^{-1/2}$ was uncovered, where $c$ is a constant that depends on the complexity of the state under consideration. We found that within the $Z_2$ phase, this constant is approximately 0.096, while at the critical point, it increases to 0.28. For fixed infidelity error $\epsilon = 1 - F$, this scaling implies that the number of required training samples $|\mathcal{T}|$ is quadratic in the number of qubits $N$. Note that this is for fixed bond dimension, such that the overall number of model parameters $N_p$ scales only linearly in $N$. Similar scaling has been observed for RBM tomography schemes, except the RBM tomography schemes require a quadratic number of parameters with a linear number of training samples to achieve a fixed error $\epsilon$ for Ising ground states, swapping the BEBM scaling for $N_p$ and $|\mathcal{T}|$ \cite{sehayek_learnability_2019, torlai_integrating_2019}. 


\section{Discussion and Conclusion}

The characterization of quantum many body states lies at the heart of quantum computing and quantum information. This task becomes even more crucial as we rapidly progress toward quantum systems of larger size. In this paper, we have presented an alternative machine learning method that harnesses both the power of generative models and the probabilistic nature of quantum mechanics for pure quantum state tomography. The scheme learns from a minimal number of probability distributions obtained from easily accessible quantum measurements such as single Pauli measurements. 

While describing a generic quantum many body state requires an exponential number of parameters, quantum states that occur in many practical situations possess features such as locality and symmetries which lead to a reduction in the required parameters \cite{Poulin2011}. In particular, inspired by the fact that the internal architecture of a tensor network provides an efficient method for targeting states that are concentrated in some part of the Hilbert space, we have utilized the Born machine for learning across different quantum ordered phases. We have further shown that extending the tensor-based Born machine to complex values and encoding data from non-native measurement bases (basis-enhancement) not only leads to an increase in model expressibilty, but also allows one to access information about the quantum relative phases, which is essential for full reconstruction of non-trivial pure quantum states. While our model can be generalized to any tensor network architecture, we have utilized the one dimensional matrix product state as a model ansatz for our basis-enhanced Born machine, and we have shown that it is capable of learning across the emergent $Z_2$, $Z_3$ and $Z_4$ ordered phases of a Rydberg system with fidelities surpassing 99\%. Surprisingly, our scheme is even capable of reconstructing quantum states at and close to critical points, which tend to be much harder to learn due to the existence of long range correlations. 

If enough samples are gathered, information from a standard basis such as $z$ would theoretically be sufficient to reconstruct quantum states that have real and non-nonegative amplitudes (which generally occurs for the ground states of stoquastic Hamiltonians \cite{bravyi_complexity_2007}). While our numerical results for the XY Hamiltonian confirm that learning only from the $z$-basis is sufficient to reconstruct the ground state in many parts of the phase diagram, in the oscillatory phase, the $z$-basis probability distribution is spread across a larger fraction of the $2^N$ possible configurations $\mathbf{v}$. This complex structure poses an obstacle to learning the single-basis probability distribution, but it is still possible to reconstruct such ground states via basis-enhancement.

More precisely, the presence of barren plateaus in the optimization landscape may explain the single-basis Born machine's failure in learning the ground state of the XY chain in the oscillatory region. We found that the probability amplitudes of this particular state were qualitatively identical to the other XY chain ground states we examined, so one might expect the single-basis Born machine to capture this state as it did for the other points. We believe that the large Shannon entropy associated with the $z$-basis data for this state may lead to a barren plateau -- a region in parameter space where the gradient vanishes, halting optimization \cite{mcclean_barren_2018}. Barren plateaus have been observed in tensor network training scenarios, and it is likely that the global nature of the NLL (Eq. (\ref{eq:OG Loss})) contributes to their presence in the Born machine's training landscape \cite{liu_presence_2021}. Even so, the basis-enhanced Born machine can capture the oscillatory ground state of the XY chain. Providing the $x$-basis training data -- which has much lower Shannon entropy -- in addition to the $z$-basis data improves optimization while still providing the necessary state information. Using complex parameters greatly increases the number of identical solutions up to an overall phase, such that locating one of the global minimums is much more likely regardless of model initialization.

We have shown that complete ground state reconstruction is achievable using only two Pauli measurement bases if care is taken to select these bases using a priori knowledge of which operators appear in the Hamiltonian. However, when no a priori knowledge about the state or Hamiltonian is available, it may still be possible to use the BEBM scheme to achieve quantum state reconstruction. The naive approach would be to include data from three orthogonal bases ($x$, $y$, and $z$) in training, although this may include redundant information. Another approach would involve using information about the training data sets' Shannon entropies for basis selection, either prior to learning  or adaptively as in \cite{lange_adaptive_2022}. A few global bases that are experimentally easily accessible could be considered as candidates, and the Shannon entropy of each could be estimated from a small amount of data. The training bases should be selected such the net Shannon entropy is small, thereby increasing the information provided to the model but also improving the trainability of the Born machine. If only one basis possesses a small Shannon entropy, a secondary basis could be selected that is rotated only partially ($<\pi/2$) from the first. We also note that more measurement bases and either a matrix product operator architecture or a purification scheme such as that presented in \cite{torlai_latent_2018} will generally be necessary for mixed state reconstruction, but further study is required. Moreover, methods beyond stochastic gradient descent for optimization of tensor networks should be explored to relax the importance of training data basis choice. 


The basis-enhanced tensor-based Born machine is a useful tool for quantum state reconstruction in the current age of noisy devices and limited experimentally accessible measurements. The scheme could additionally be used to compute various non-local quantities that are hard to measure, including von Neumann entanglement entropy, higher correlation functions, or even local observables in a particular basis \cite{Omran_2019, ebadi2021quantum}. While in this paper we have focused on reconstructing quantum states across phases with well-defined symmetries, the BEBM could be applied to learning less trivial phases, such as the recently realized topological phases or spin liquids \cite{Semeghini_2021}. Applying the BEBM to other non-local states like the GHZ state or cluster states that are challenging to learn using a single basis Born machine would be another important future direction \cite{Najafi_GHZ_nonlocal}.




\begin{acknowledgments}
K.N would likes to thank Xun Gao, Mohammad Ali Rajabpour, Anatoli Dymarsky and Dolev Bluvstein for their helpful discussions. A.M.G acknowledges support from the NSF through the Graduate Research Fellowships Program, as well as support through the Theodore H. Ashford Fellowships in the Sciences. S.F.Y would like to acknowledge funding by the NSF and DOE.

\end{acknowledgments}

\bibliography{main}

\appendix

\section{Supplementary Material}
\subsection{Characterization of the 1D Rydberg Phase Diagram}
\subsubsection{Locating a Critical Point} 
\begin{figure*}
\centering
    \begin{subfigure}[b]{0.49\textwidth}
        \centering
        \includegraphics[width = 8.7cm]{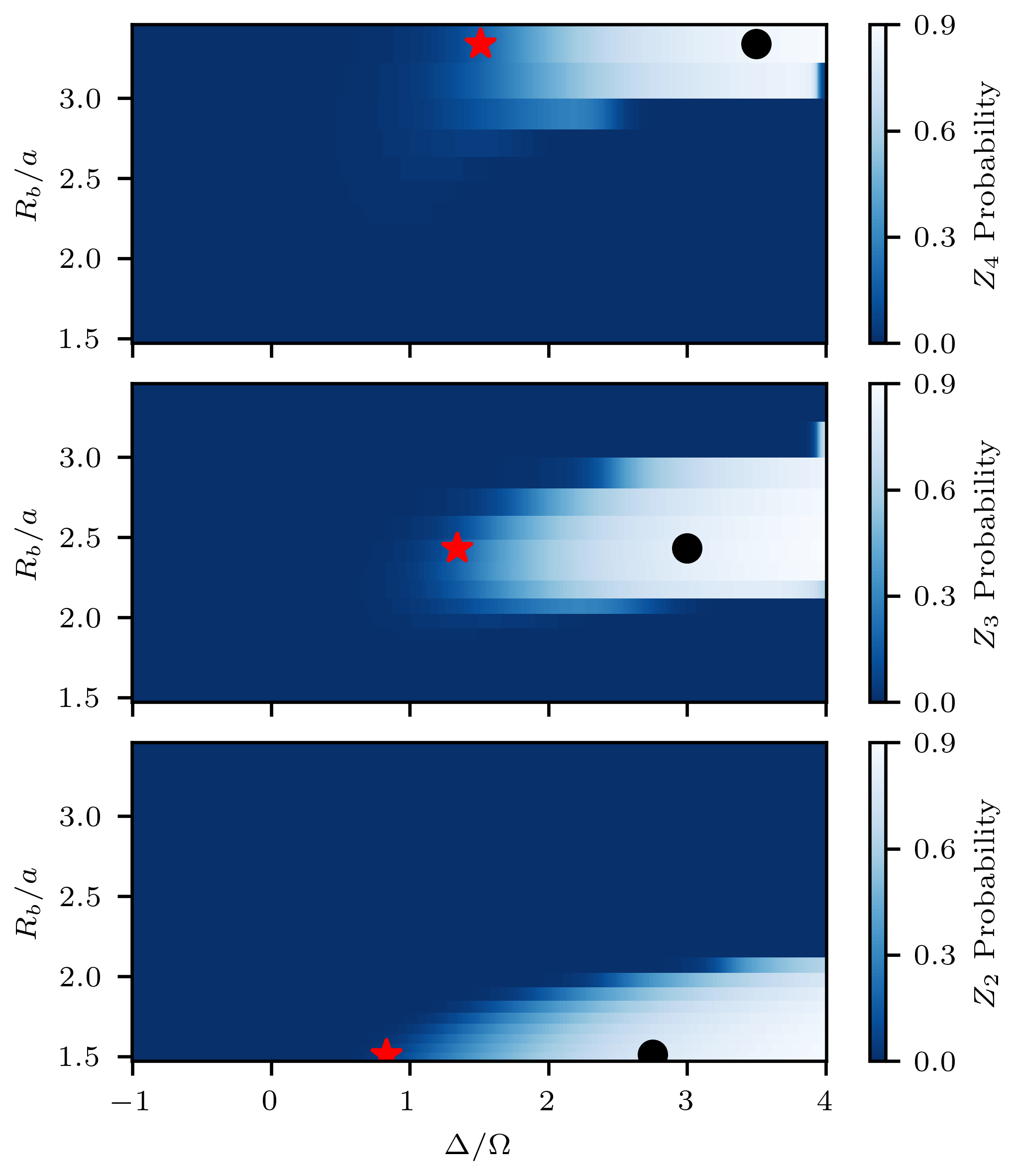}
        \caption{}
        \label{subfig:phase-map}
    \end{subfigure}
    \hfill
    \begin{subfigure}[b]{0.49\textwidth}
        \centering
        \includegraphics[width = 8.7cm]{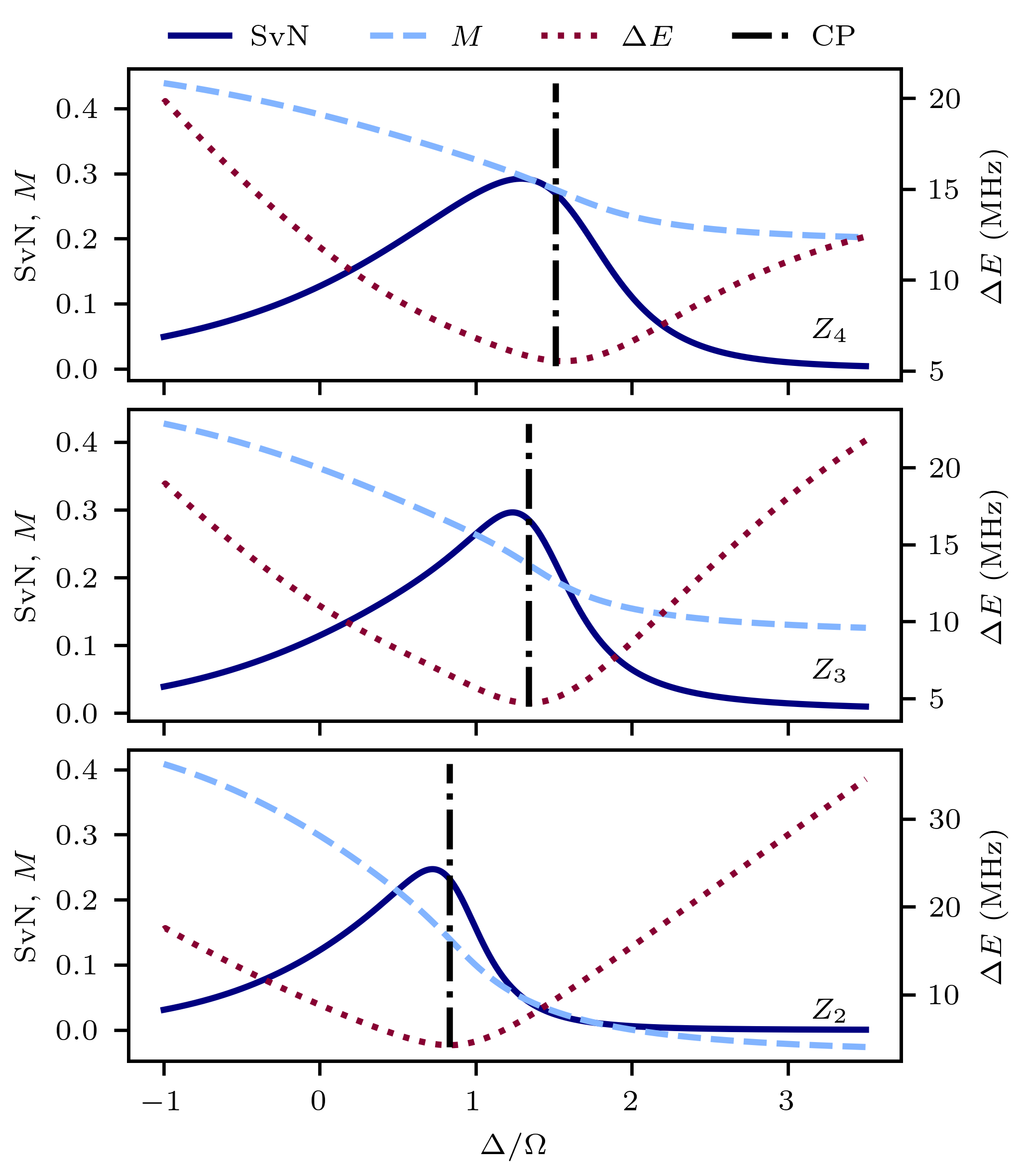}
        \caption{}
        \label{subfig:crit-pts}
    \end{subfigure}
    
    \begin{subfigure}[b]{1.00\textwidth}
        \centering
        \includegraphics[width = 17.8cm]{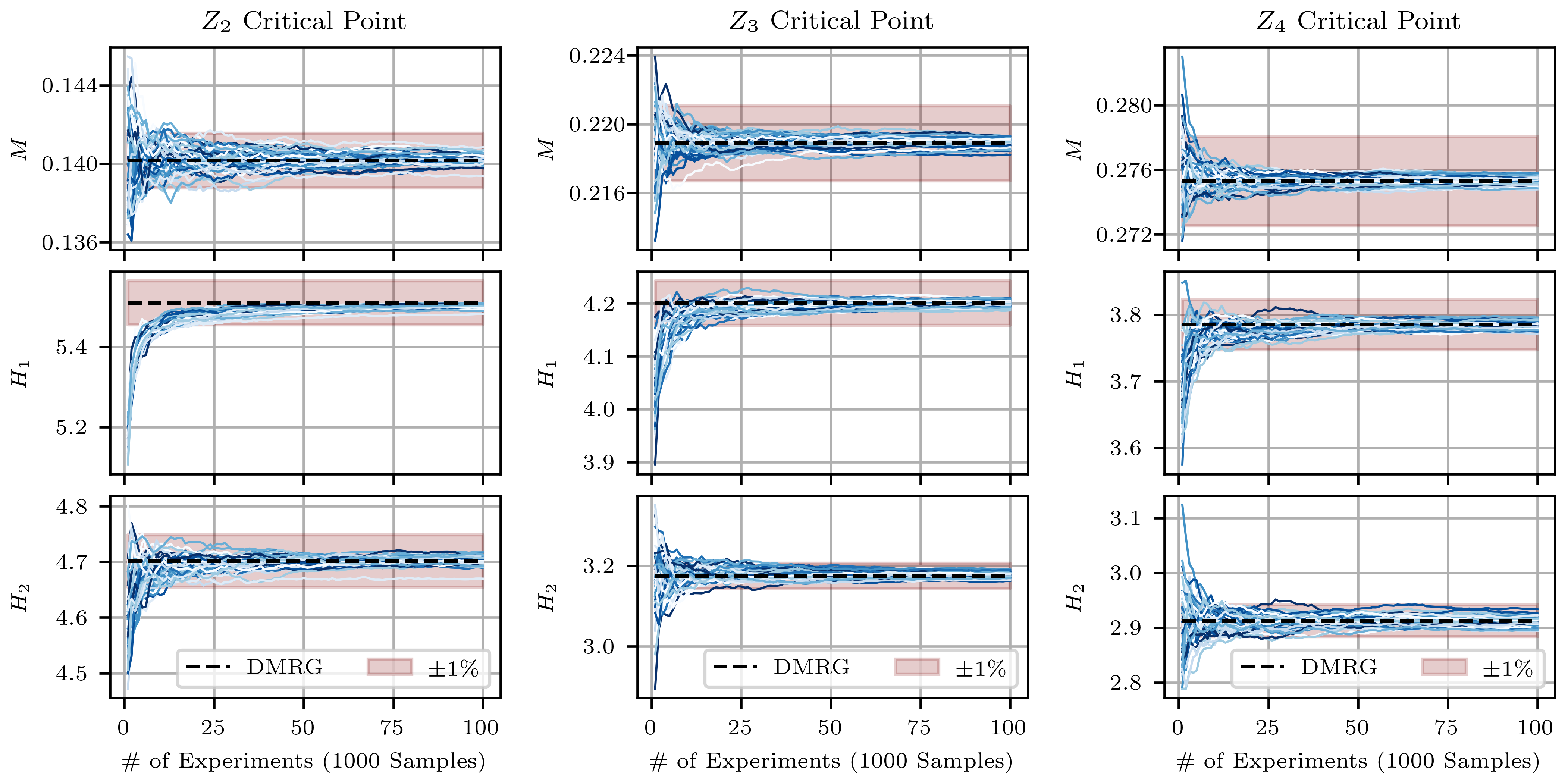}
        \caption{}
        \label{subfig:MC-sampling}
    \end{subfigure}
\caption{\ref{subfig:phase-map} Numerical phase diagram of a 1D Rydberg chain with 13 atoms. Red stars indicate critical points between ordered phases and disorder. Black dots indicate points in phase selected for study. \ref{subfig:crit-pts} Observables examined along fixed $R_b/a$ to determine critical point location. Located critical points are indicated by the vertical black dot-dashed lines. \ref{subfig:MC-sampling} Magnetization ($M$), first order Rényi entropy ($H_1$), and second order Rényi entropy ($H_2$) results from 50 unique sampling trajectories. Observables are calculated cumulatively from the samples over the course of one trajectory.}
\end{figure*}

A numerical phase diagram for a 1D Rydberg chain of 13 atoms is provided in Fig. \ref{subfig:phase-map}. To produce this figure, the ground state at different $a$ and $\Delta$ (for fixed $R_b$ and $\Omega$) was calculated using DMRG. The double sum appearing in the Hamiltonian was truncated to include at most 5 of the nearest neighbors. The probabilities indicated in the figure by the color bar were found by calculating the overlap of the ground state MPS with the corresponding ordered product state. 

Three different observables were examined along fixed $R_b/a$ lines to locate the critical points: the bipartite entanglement entropy $\mathrm{SvN}$, the magnetization $M$, and the first excited energy gap $\Delta E$. The bipartite entanglement entropy approaches a maximum near the critical point due to diverging correlations. The slope of the magnetization will approach an extremum at the critical point as the preferential ground state shifts from all spins aligned to an ordered crystalline phase; this is less noticeable for the higher order phases, as fewer spins are flipped within the ordered phase. Finally, for our choice of $N = 13$, the first excited energy gap will approach a minimum at the critical point. The energy gap would vanish throughout the ordered phase for the case of an infinite chain due to the degenerate cyclic permutations of the ordered state; however, when the size of the unit cell within the ordered phase and the length of a finite chain are congruent modulo 1, there is a preference for one specific configuration of the ordered phase over any otherwise degenerate permutation. In this special configuration, the atoms on either end of the chain (which have only one neighbor and therefore a smaller interaction energy) are excited to the Rydberg state, resulting in a lower energy for a finite chain. For a chain of 13 atoms, the $Z_2$, $Z_3$, and $Z_4$ ordered phases all have units cells that are congruent modulo 1 with the chain length. This breaks the degeneracy and the energy gap increases after the phase boundary is crossed in Fig. \ref{subfig:crit-pts} for this reason.

\subsubsection{Monte Carlo Sampling}
Prior to uncovering the scaling law discuessed in the main text, we used a Monte Carlo approach to estimate the number of samples $|\mathcal{T}|$ required to accurately capture a quantum state of interest. Two observables -- magnetization ($M = (1/N)\sum_{i} \langle S_{z}^{i} \rangle$) and Renyi entropy ($H_{\alpha}(\mathbbm{P}) = (1 - \alpha)^{-1}\ln{\sum_{\mathbf{v} \in \mathcal{T}} \mathbbm{P}(\mathbf{v})^{\alpha}}$) -- were considered, as these can easily be quantified from a discrete number of samples while providing insight into the quantum state. 

The system was sampled 100,000 times in the computational basis, and these samples were cumulatively used to estimate the expectation value of the system's magnetization and Renyi entropy  every 1,000 samples, creating a single ``sampling trajectory.'' This process was repeated 50 times. The number of samples required to capture the state was estimated by determining when the sampling trajectories converged within $\pm$1\% of the true target observable, which was calculated from the DMRG output. After reviewing the Monte Carlo results (see Fig. \ref{subfig:MC-sampling}), 30,000 samples were used to form each training data set.

\subsection{Additional Results}
\subsubsection{1D Rydberg Chain}
\begin{figure*}
\centering
    \begin{subfigure}[b]{1.0\textwidth}
        \centering
        \includegraphics[width = 17.8cm]{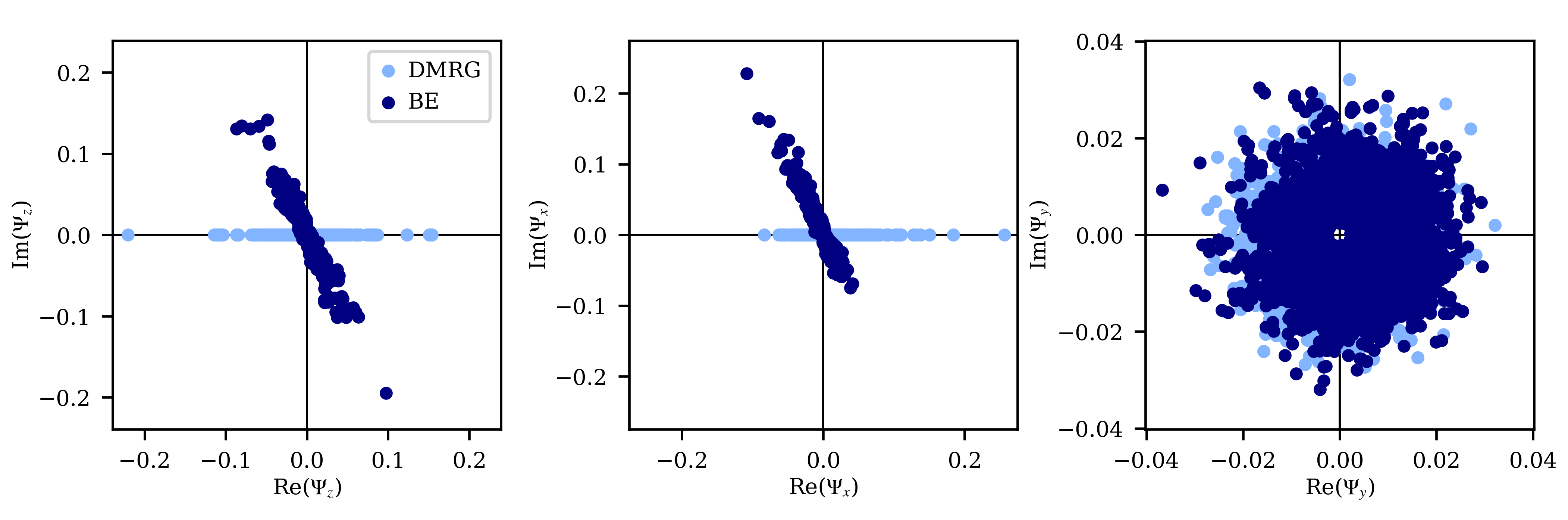}
        \caption{}
        \label{subfig:amps-Z2CP}
    \end{subfigure}
    
    \begin{subfigure}[b]{1.0\textwidth}
        \centering
        \includegraphics[width = 17.8cm]{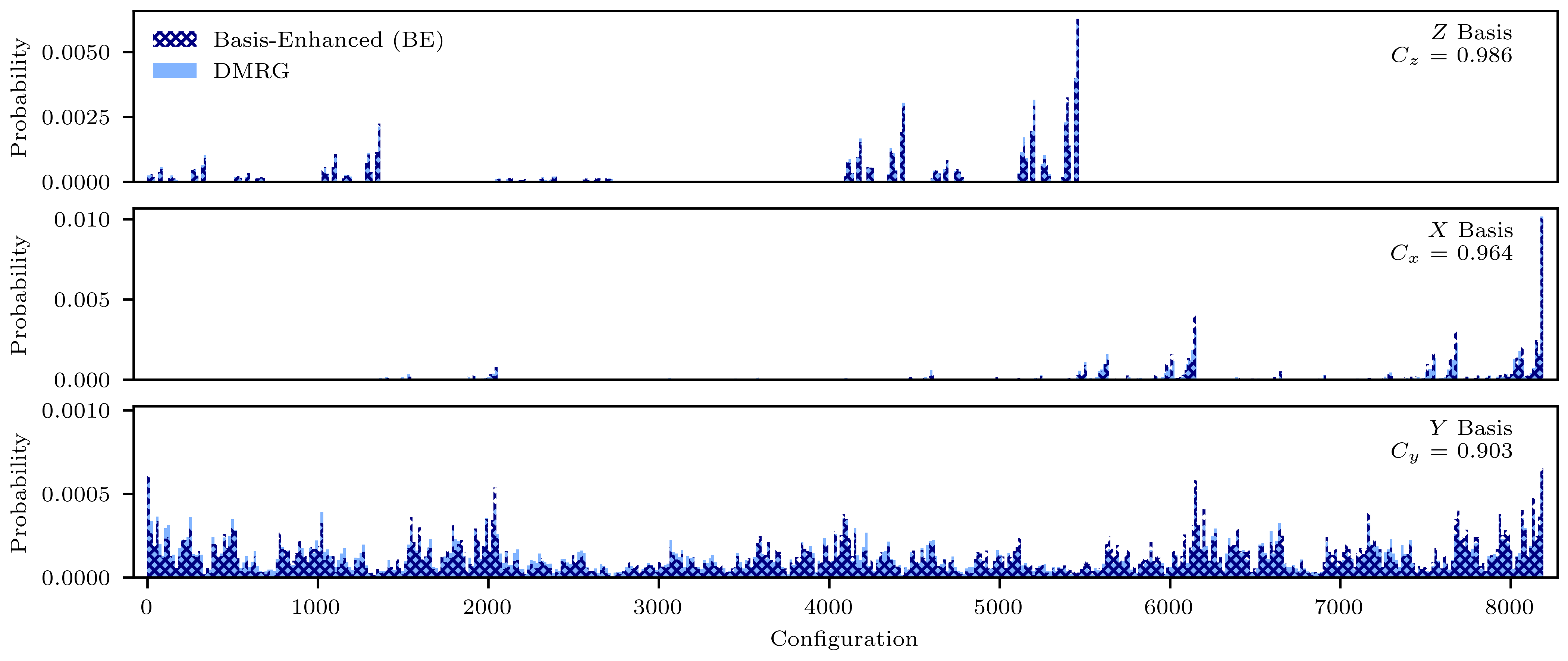}
        \caption{}
        \label{subfig:hists-Z2CP}
    \end{subfigure}
\caption{\ref{subfig:amps-Z2CP} Probability amplitudes in the $z$-, $x$-, and $y$-bases calculated from the DMRG MPS and from the trained basis-enhanced Born machine at the $Z_2$ critical point. \ref{subfig:hists-Z2CP} Overlaid histograms of the probability distributions in the $z$-, $x$-, and $y$-bases from the DMRG MPS and from the trained basis-enhanced Born machine at the $Z_2$ critical point.}
\label{fig:Z2CP}
\end{figure*}

\begin{figure*}
\centering
    \begin{subfigure}[b]{1.0\textwidth}
        \centering
        \includegraphics[width = 17.8cm]{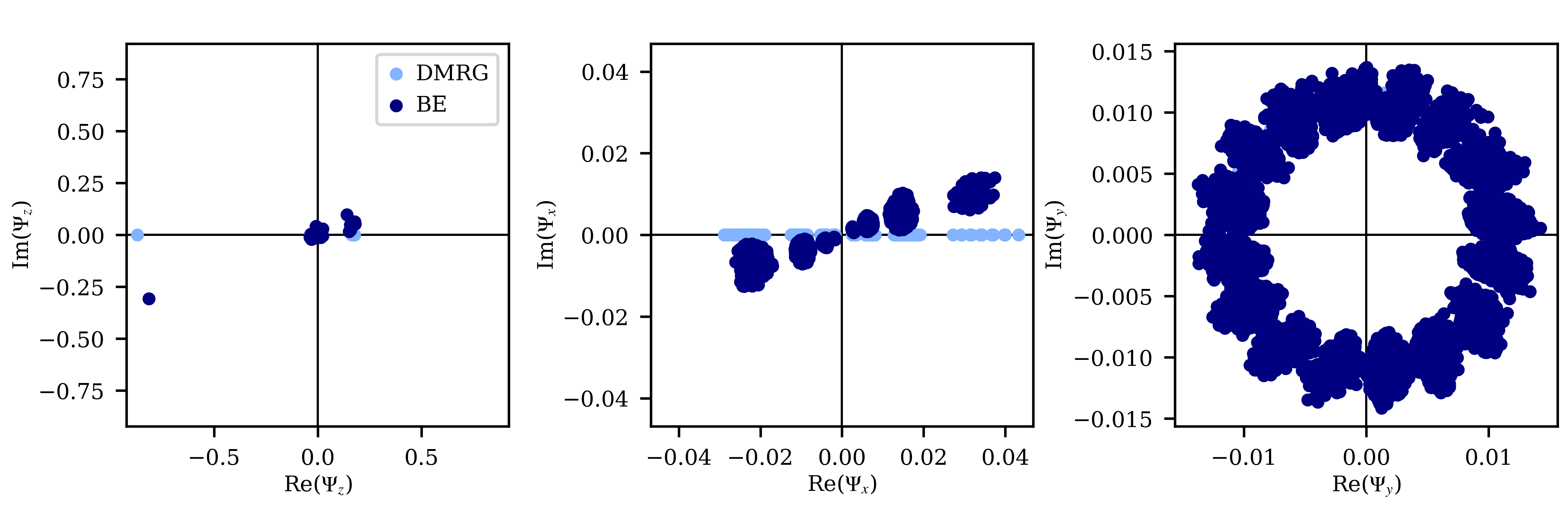}
        \caption{}
        \label{subfig:amps-Z2PH}
    \end{subfigure}
    
    \begin{subfigure}[b]{1.0\textwidth}
        \centering
        \includegraphics[width = 17.8cm]{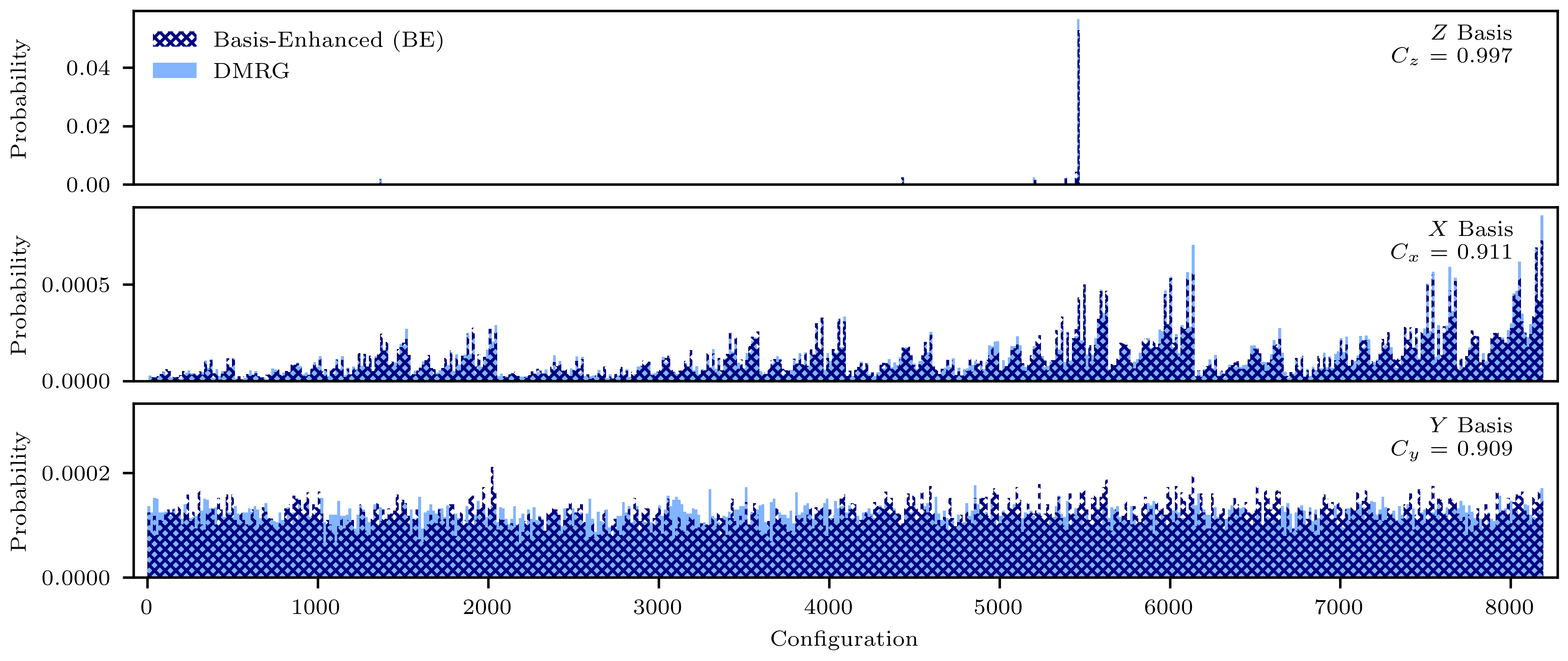}
        \caption{}
        \label{subfig:hists-Z2PH}
    \end{subfigure}
\caption{\ref{subfig:amps-Z2PH} Probability amplitudes in the $z$-, $x$-, and $y$-bases calculated from the DMRG MPS and from the trained basis-enhanced Born machine within the $Z_2$ ordered phase. \ref{subfig:hists-Z2PH} Overlaid histograms of the probability distributions in the $z$-, $x$-, and $y$-bases from the DMRG MPS and from the trained basis-enhanced Born machine within the $Z_2$ ordered phase.}
\label{fig:Z2PH}
\end{figure*}

\begin{figure*}
\centering
    \begin{subfigure}[b]{1.0\textwidth}
        \centering
        \includegraphics[width = 17.8cm]{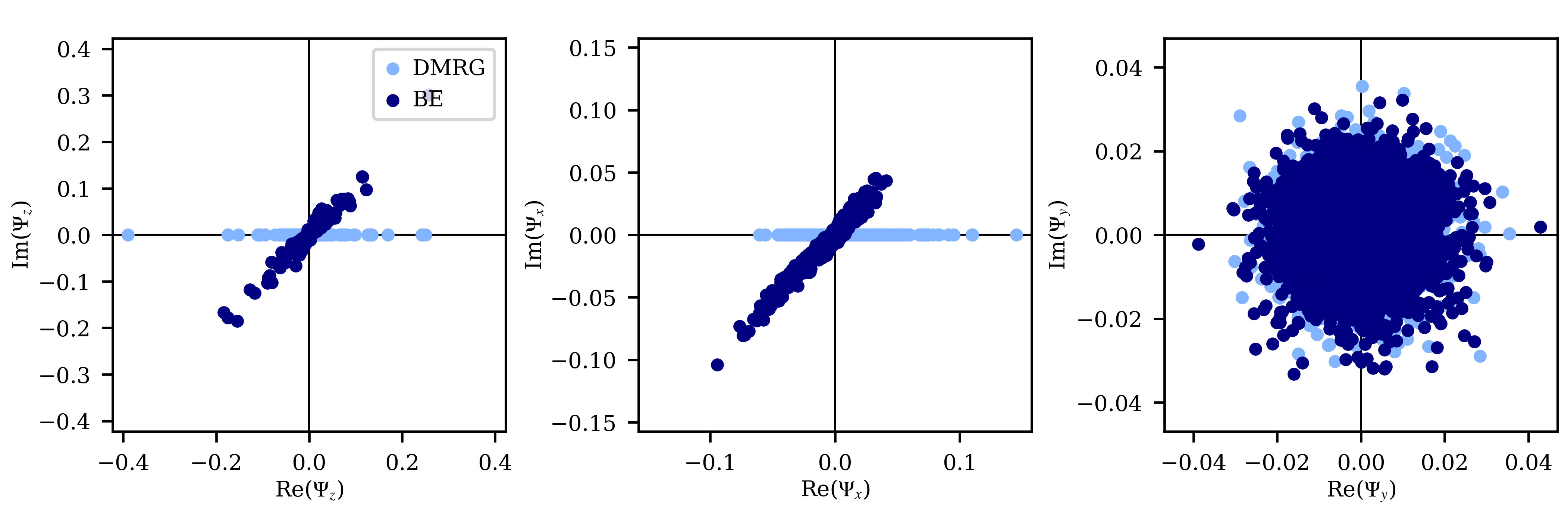}
        \caption{}
        \label{subfig:amps-Z3CP}
    \end{subfigure}
    
    \begin{subfigure}[b]{1.0\textwidth}
        \centering
        \includegraphics[width = 17.8cm]{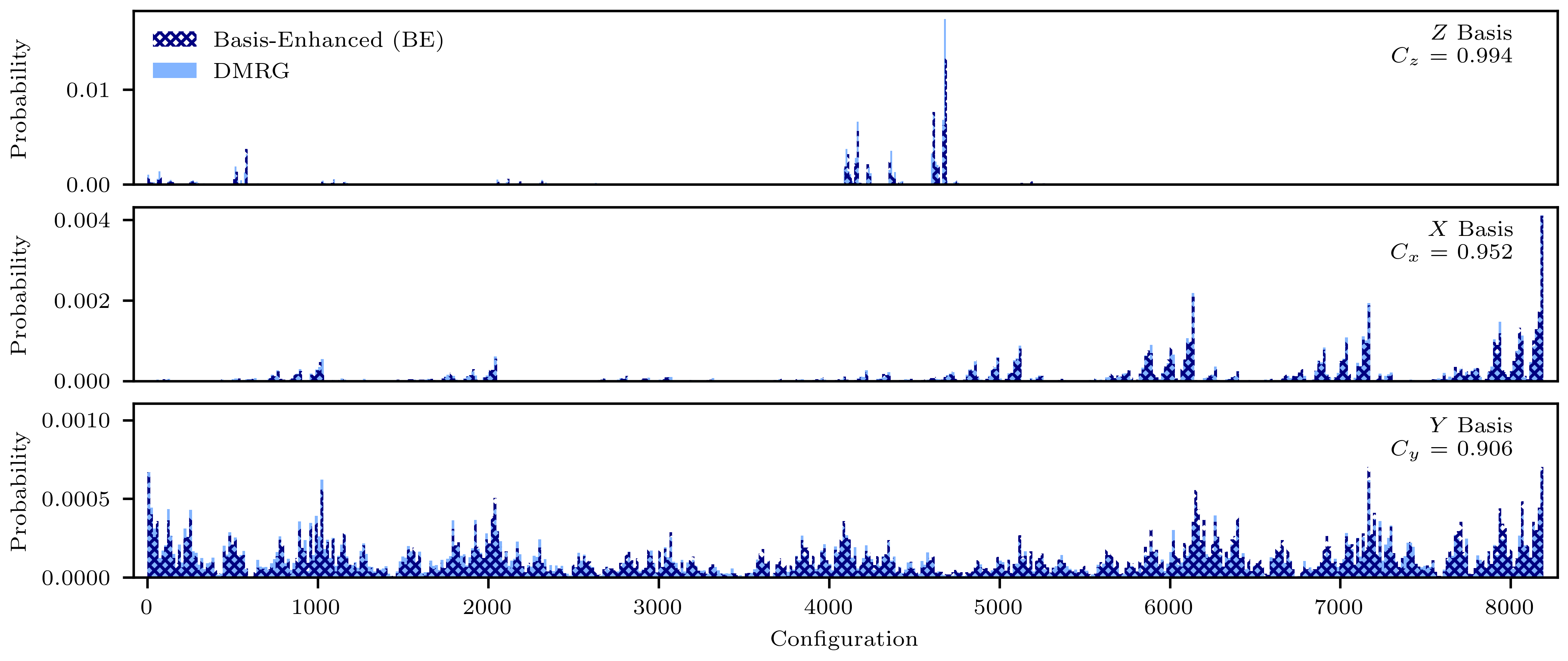}
        \caption{}
        \label{subfig:hists-Z3CP}
    \end{subfigure}
\caption{\ref{subfig:amps-Z3CP} Probability amplitudes in the $z$-, $x$-, and $y$-bases calculated from the DMRG MPS and from the trained basis-enhanced Born machine at the $Z_3$ critical point. \ref{subfig:hists-Z3CP} Overlaid histograms of the probability distributions in the $z$-, $x$-, and $y$-bases from the DMRG MPS and from the trained basis-enhanced Born machine at the $Z_3$ critical point.}\label{fig:Z3CP}
\end{figure*}

\begin{figure*}
\centering
    \begin{subfigure}[b]{1.0\textwidth}
        \centering
        \includegraphics[width = 17.8cm]{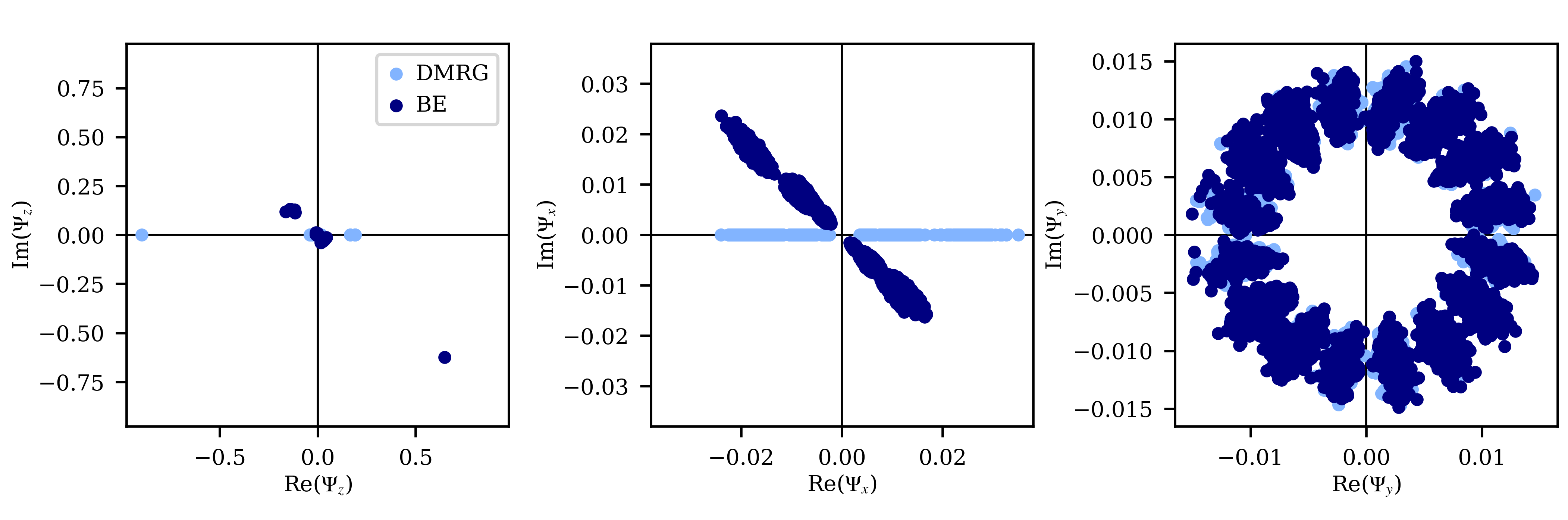}
        \caption{}
        \label{subfig:amps-Z3PH}
    \end{subfigure}
    
    \begin{subfigure}[b]{1.0\textwidth}
        \centering
        \includegraphics[width = 17.8cm]{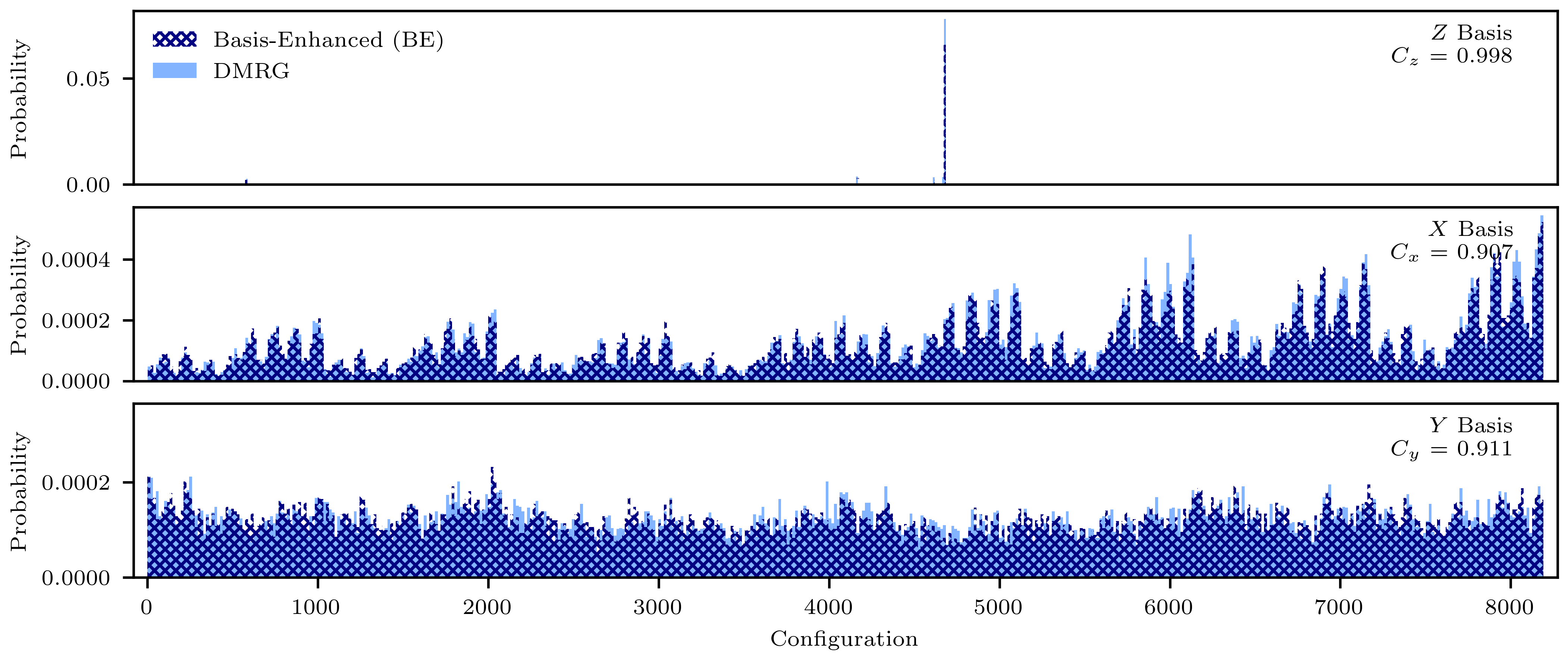}
        \caption{}
        \label{subfig:hists-Z3PH}
    \end{subfigure}

\caption{\ref{subfig:amps-Z3PH} Probability amplitudes in the $z$-, $x$-, and $y$-bases calculated from the DMRG MPS and from the trained basis-enhanced Born machine within the $Z_3$ ordered phase. \ref{subfig:hists-Z3PH} Overlaid histograms of the probability distributions in the $z$-, $x$-, and $y$-bases from the DMRG MPS and from the trained basis-enhanced Born machine within the $Z_3$ ordered phase.}
\label{fig:Z3PH}
\end{figure*}

\begin{figure*}
\centering
    \begin{subfigure}[b]{1.0\textwidth}
        \centering
        \includegraphics[width = 17.8cm]{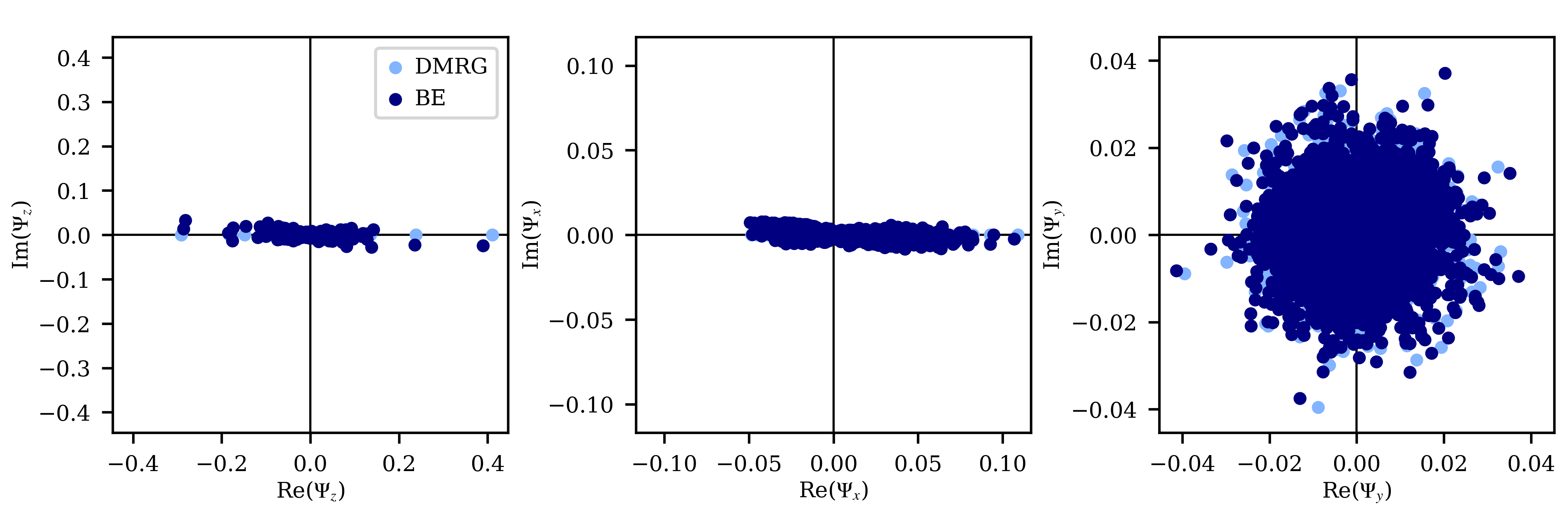}
        \caption{}
        \label{subfig:amps-Z4CP}
    \end{subfigure}
    
    \begin{subfigure}[b]{1.0\textwidth}
        \centering
        \includegraphics[width = 17.8cm]{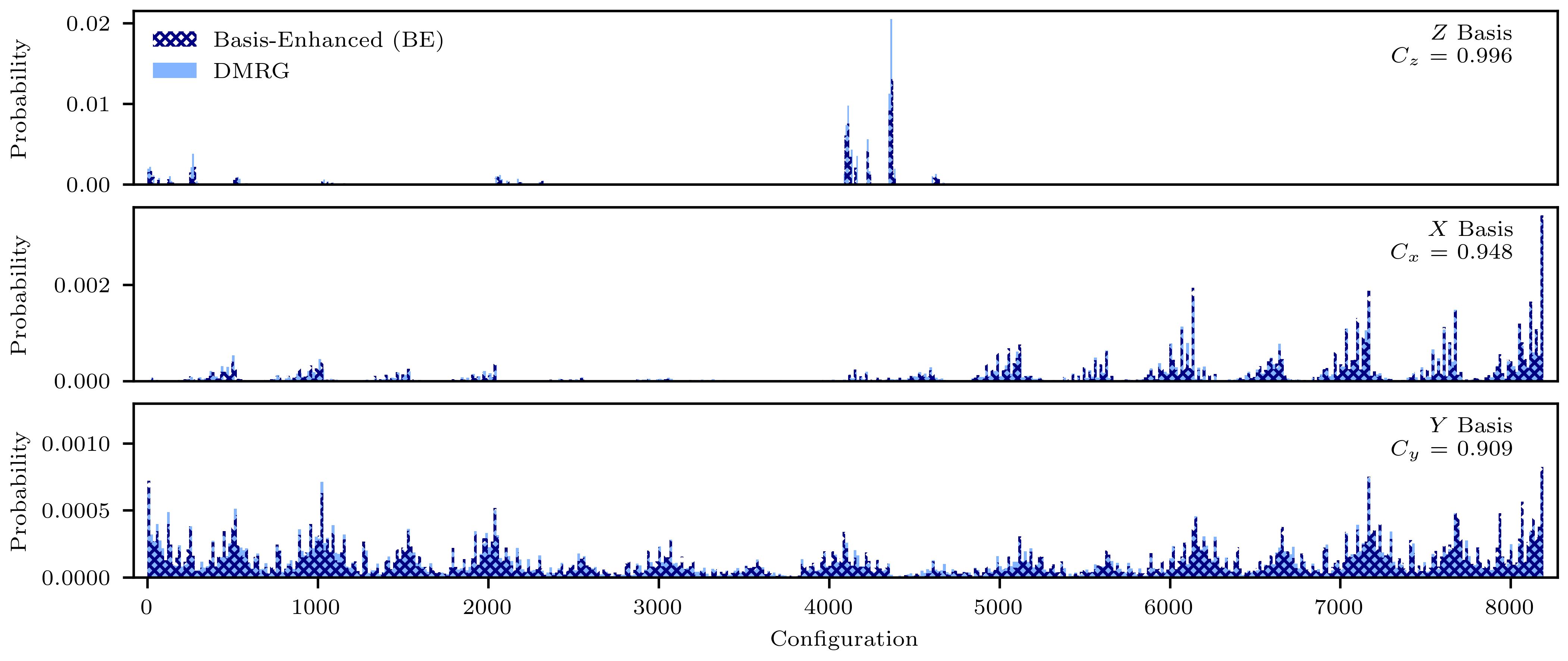}
        \caption{}
        \label{subfig:hists-Z4CP}
    \end{subfigure}
\caption{\ref{subfig:amps-Z4CP} Probability amplitudes in the $z$-, $x$-, and $y$-bases calculated from the DMRG MPS and from the trained basis-enhanced Born machine at the $Z_4$ critical point. \ref{subfig:hists-Z4CP} Overlaid histograms of the probability distributions in the $z$-, $x$-, and $y$-bases from the DMRG MPS and from the trained basis-enhanced Born machine at the $Z_4$ critical point.}
\label{fig:Z4CP}
\end{figure*}

\begin{figure*}
\centering
    \begin{subfigure}[b]{1.0\textwidth}
        \centering
        \includegraphics[width = 17.8cm]{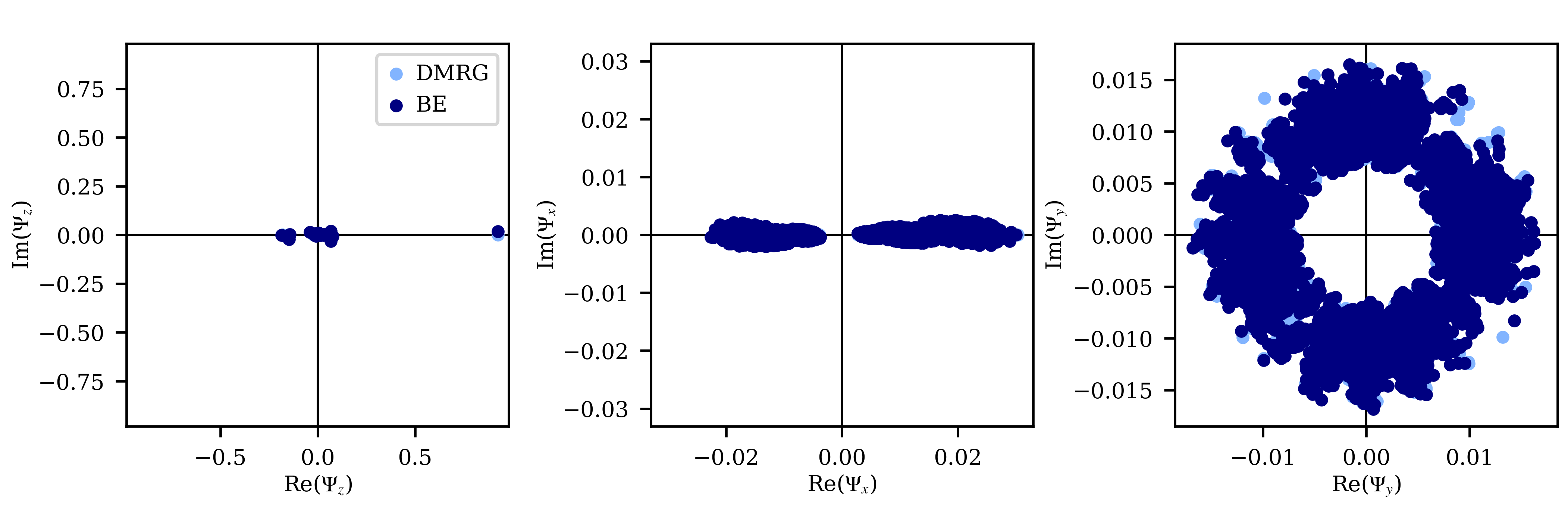}
        \caption{}
        \label{subfig:amps-Z4PH}
    \end{subfigure}
    
    \begin{subfigure}[b]{1.0\textwidth}
        \centering
        \includegraphics[width = 17.8cm]{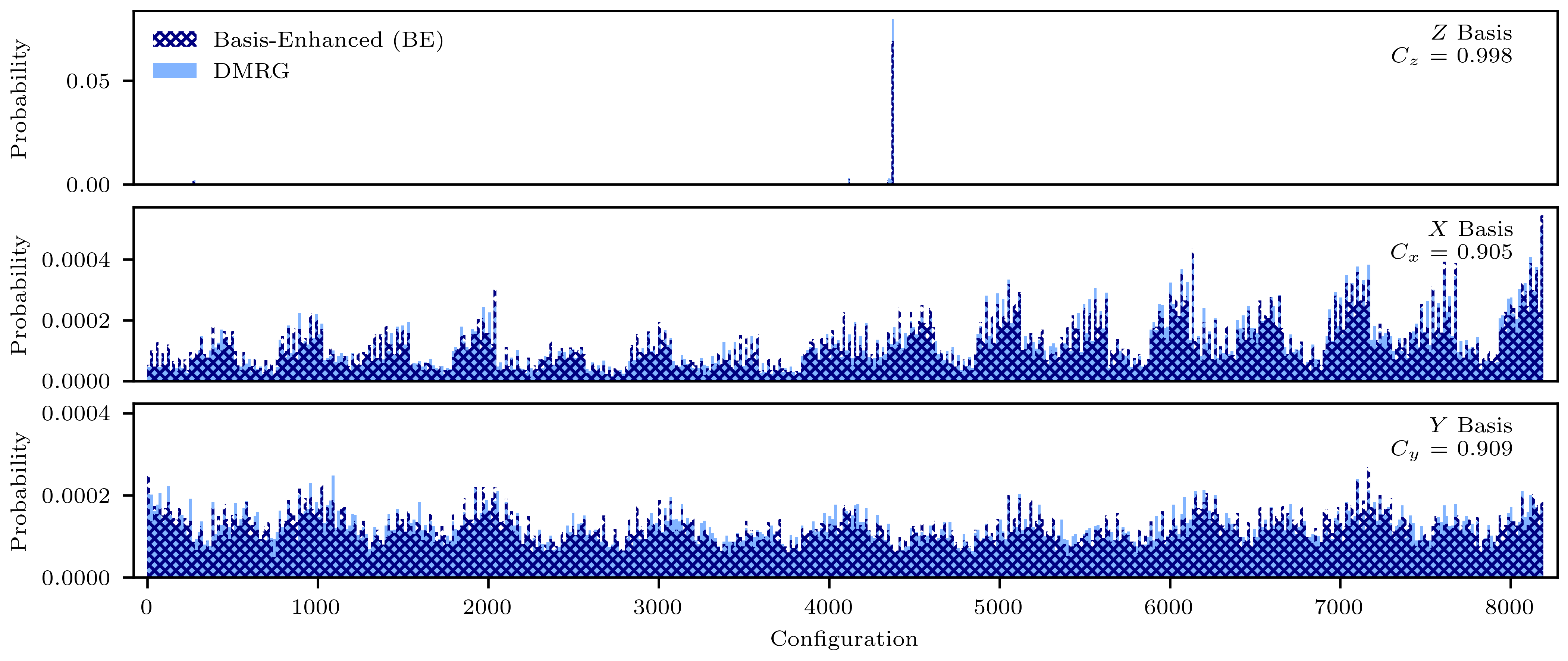}
        \caption{}
        \label{subfig:hists-Z4PH}
    \end{subfigure}
\caption{\ref{subfig:amps-Z4PH} Probability amplitudes in the $z$-, $x$-, and $y$-bases calculated from the DMRG MPS and from the trained basis-enhanced Born machine within the $Z_4$ ordered phase. \ref{subfig:hists-Z4PH} Overlaid histograms of the probability distributions in the $z$-, $x$-, and $y$-bases from the DMRG MPS and from the trained basis-enhanced Born machine within the $Z_4$ ordered phase.}
\label{fig:Z4PH}
\end{figure*}

Additional results for points across the 1D Rydberg chain phase diagram are provided in Fig. \ref{fig:Z2CP} - \ref{fig:Z4PH}. 

\subsubsection{1D Anisotropic XY Chain}
\begin{figure*}
\centering
    \begin{subfigure}[b]{1.0\textwidth}
        \centering
        \includegraphics[width = 17.8cm]{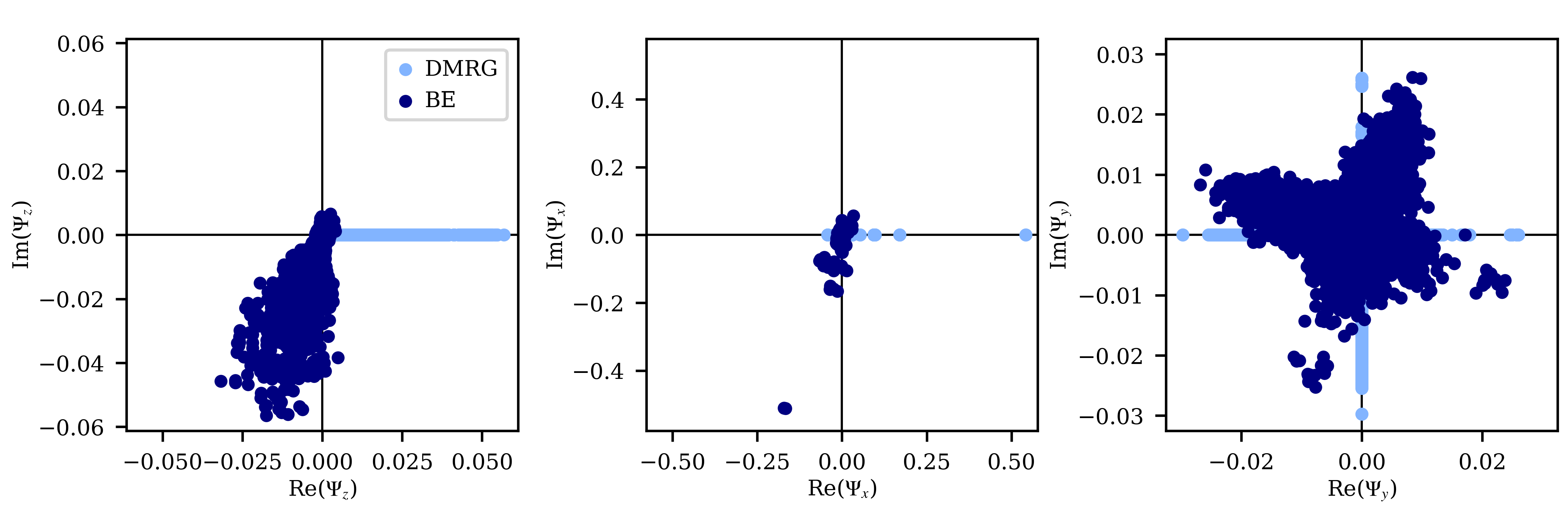}
        \caption{}
        \label{subfig:amps-oscillatory}
    \end{subfigure}

    \begin{subfigure}[b]{1.0\textwidth}
        \centering
        \includegraphics[width = 17.8cm]{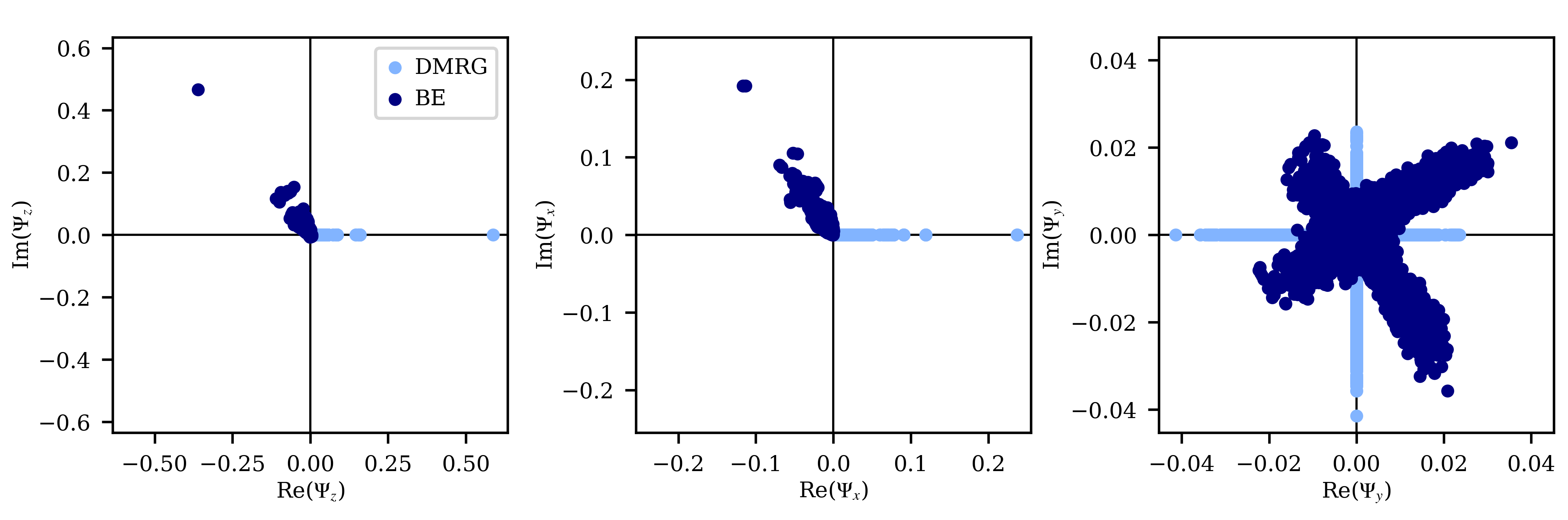}
        \caption{}
        \label{subfig:amps-isingCP}
    \end{subfigure}
    
    \begin{subfigure}[b]{1.0\textwidth}
        \centering
        \includegraphics[width = 17.8cm]{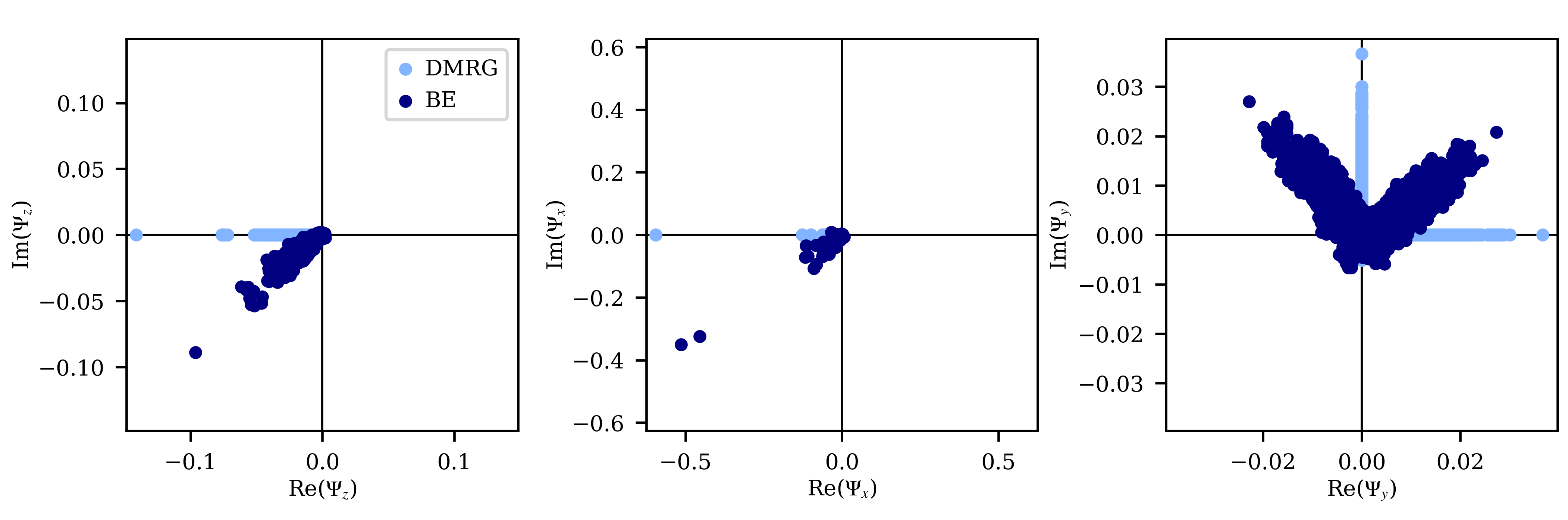}
        \caption{}
        \label{subfig:amps-ordered}
    \end{subfigure}
\caption{\ref{subfig:amps-oscillatory} Probability amplitudes in the $z$-, $x$-, and $y$-bases calculated from the DMRG MPS and from the trained basis-enhanced Born machine at a point from the oscillatory region of the 1D anistropic XY phase diagram. \ref{subfig:amps-isingCP} Probability amplitudes in the $z$-, $x$-, and $y$-bases calculated from the DMRG MPS and from the trained basis-enhanced Born machine at the Ising critical point of the 1D anistropic XY phase diagram. \ref{subfig:amps-ordered} Probability amplitudes in the $z$-, $x$-, and $y$-bases calculated from the DMRG MPS and from the trained basis-enhanced Born machine at a point from the ordered region of the 1D anistropic XY phase diagram.}
\label{fig:XY}
\end{figure*}

More in-depth results across the 1D anisotropic XY chain phase diagram are provided in Fig. \ref{fig:XY}.

\subsubsection{Testing Basis Dependence}
\begin{figure}
    \centering
    \includegraphics{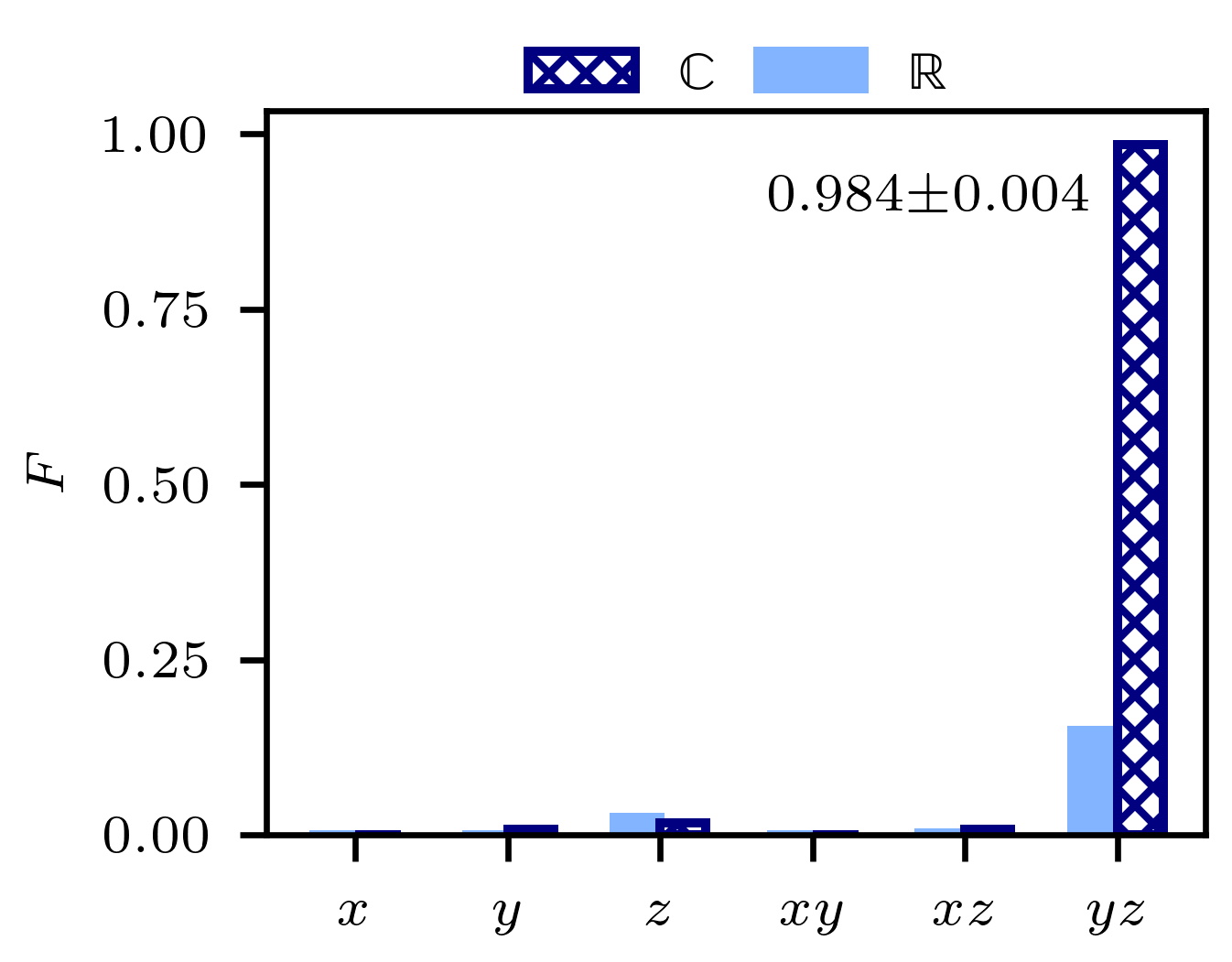}
    \caption{Final quantum fidelity after training a Born machine with different combinations of training data basis (indicated by the $x$-axis) and model parameters (real or complex) at the $Z_2$ critical point of the altered Rydberg Hamiltonian. }
    \label{fig:fidbar_yRydChain}
\end{figure}

To test the idea that the basis-dependent success of basis-enhanced Born machine can be partially explained by the most prominent bases of operators appearing in the Hamiltonian, we constructed an altered Rydberg chain Hamiltonian in which the $\sigma_x$ operator was replaced by a $\sigma_y$ operator. All combinations of training data basis(es) and model parameters were tested for the $Z_2$ critical point of this altered Hamiltonian; the results are provided in Fig. \ref{fig:fidbar_yRydChain}. The basis-enhanced Born machine is successful for the choice of $yz$ learning and complex model parameters, in accordance with the basis of the operators appearing in the altered Hamiltonian.

\end{document}